\documentclass[12pt]{article}

\usepackage[margin=1in]{geometry}

\usepackage[authoryear,longnamesfirst]{natbib}

\usepackage{graphicx}
\usepackage[english]{babel}
\usepackage{array}
\usepackage{float}
\usepackage{placeins}
\usepackage{amsmath}
\usepackage{amssymb}
\usepackage{multirow}
\usepackage{booktabs}
\usepackage{subcaption}
\usepackage{xcolor}
\usepackage{xspace}
\usepackage{authblk}
\usepackage[colorlinks=true, linkcolor=blue, citecolor=blue, urlcolor=blue]{hyperref}

\def\tsc#1{\csdef{#1}{\textsc{\lowercase{#1}}\xspace}}
\tsc{WGM}
\tsc{QE}
\tsc{EP}
\tsc{PMS}
\tsc{BEC}
\tsc{DE}

\title{\textbf{MoTIF: A Mode-Structured Tensor Framework for\\
Multi-Parametric Approximation, Super-Resolution and\\
Forecasting of Unsteady Systems}}

\author[1]{G. Barrag\'{a}n\thanks{Corresponding author: \texttt{guillermo.barragan@alumnos.upm.es}}}
\author[1]{A. Hetherington}
\author[1]{A. Sengupta}
\author[1]{R.Abad\'{i}a-Heredia}
\author[1,2]{J. Garicano-Mena\thanks{\texttt{jesus.garicano.mena@upm.es}}}
\author[1,2]{S. Le Clainche\thanks{\texttt{soledad.leclainche@upm.es}}}

\affil[1]{\small ETSI Aeron\'{a}utica y del Espacio, Universidad Polit\'{e}cnica de Madrid,
Plaza Cardenal Cisneros, 3, 28040 Madrid, Spain}
\affil[2]{\small Center for Computational Simulation (CCS), Campus de Montegancedo,
28660 Boadilla del Monte, Spain}

\date{}

\begin{document}

\maketitle

\begin{abstract}
We introduce MoTIF, a mode-structured tensor framework for multi-parametric approximation, super-resolution, and temporal forecasting of high-dimensional unsteady systems. The methodology leverages High-Order Singular Value Decomposition (HOSVD) to obtain a structured multilinear representation of multi-dimensional datasets, separating physical parameters, spatial coordinates, and temporal evolution into distinct modal components. This decomposition enables the application of dedicated approximation operators to each mode. Gaussian Process Regression is employed to interpolate and extrapolate parametric and spatial modal matrices, enabling database completion and resolution enhancement, while recurrent neural networks are applied to the temporal mode to forecast system evolution. This decoupled operator-learning strategy preserves the intrinsic tensor structure while providing a flexible non-intrusive reduced-order modelling framework. The proposed methodology is validated on a database of unsteady laminar flow simulations with varying Reynolds numbers and angles of attack. Accurate reconstruction of unseen flow configurations and temporal prediction are achieved, with relative root mean square errors consistently below 2\% compared to high-fidelity simulations. The framework provides a scalable and mathematically structured alternative to conventional surrogate modelling approaches for high-dimensional parametric dynamical systems.
\end{abstract}


\bigskip

\section{Introduction}

The generation of accurate multi-parametric databases for high-dimensional dynamical systems remains computationally demanding when high spatial and temporal resolutions are required. In Computational Fluid Dynamics, the governing equations of a fluid flow: Navier-Stokes, conservation of mass and energy, among others, are solved through numerical models~\cite{FLETCHER2022299}. The analysis of complex fluid flow phenomena often entails performing numerical simulations under varying physical parameters, which involves high computational cost, particularly when exploring broad parametric design spaces or performing long-time integrations~\cite{MAZUMDAR2024108897}. Reduced-order modelling (ROM) techniques offer an alternative to address these CFD limitations. These methods approximate complex systems as simplified ones by finding a latent low-dimensional space to represent the full-order model (FOM), while drastically reducing computational cost~\cite{dar2024}.

Reduced-order models can be classified as intrusive or non-intrusive depending on whether they explicitly rely on the governing equations. Non-intrusive reduced-order models (niROMs), which are purely data-driven, have attracted considerable attention due to their flexibility and applicability to legacy simulation datasets~\cite{dar2024}. Singular value decomposition (SVD)~\cite{Sirovich},~\cite{Sirovich1987TurbulenceAT} and proper orthogonal decomposition (POD)~\cite{john_l__lumley_1967} have been extensively applied to extract the dominant patterns and structures of a flow, being able to accurately represent the high-dimensional phenomenon with just a few SVD/POD modes (it is worth mentioning that SVD is one of the two common methods used to compute POD modes; therefore, the terms POD and SVD are often used interchangeably in the literature).~\cite{mendoca},~\cite{huxiao},~\cite{BUOSO2022105604}.

High-order singular value decomposition (HOSVD), originally introduced by Tucker \cite{Tuck2006} and later formalized by ~\cite{DeLathauwer2000a}, is a multi-dimensional extension of SVD, capable of extracting the underlying multilinear structures associated with each of the dimensions of the dataset~\cite{MIAO2023139}. HOSVD has been successfully applied in aerodynamics~\cite{LORENTEcomp}~\cite{Lorenteetal}~\cite{Lorente2009HOSVD}, fluid dynamics~\cite{HETHERINGTON2024109217}, and reduced-order modelling~\cite{Moayyedi04072018}.. Its tensorial structure makes it particularly suitable for multi-parametric problems where spatial, temporal, and parametric dependencies coexist in a natural multidimensional representation.

Several works have combined modal decomposition with interpolation or machine learning techniques to extend reduced-order models across parametric dimensions. For instance, Lorente et al.~\cite{Lorente2009HOSVD} demonstrated the robustness of modal decomposition coupled with interpolation strategies for aerodynamic database generation. However, classical interpolation approaches may require dense sampling in parameter space to achieve accurate reconstructions. Gaussian Process Regression (GPR)~\cite{Rasmussen2006Gaussian} provides a probabilistic and flexible alternative for non-linear interpolation, and has been successfully coupled with POD-based models for digital twin development~\cite{Aversano2021}~\cite{Procacci2022}. 

In parallel, deep learning approaches have been incorporated into niROM frameworks to enhance spatial resolution and enable temporal forecasting. For example, Hetherington et al.~\cite{HETHERINGTON2024109217} and Díaz-Morales et al.~\cite{diaz_morales2024} investigated the use of modal decomposition combined with neural networks for spatial super-resolution. Abadía-Heredia et al.~\cite{ABADIAHEREDIA2022115910} and Du et al.~\cite{DU2024105651} employed POD combined with Long Short-Term Memory (LSTM) architectures. Hochreiter \&\ Schmidhuber~\cite{hochreiter1997long} for temporal prediction of fluid dynamics time series.

Despite these advances, existing approaches typically focus on a single task—either parametric interpolation, spatial super-resolution, or temporal forecasting—and often rely on monolithic surrogate models applied to reduced coordinates. Moreover, the tensorial structure of multi-parametric datasets is not systematically exploited to assign distinct approximation operators tailored to the mathematical characteristics of each dimension.

In multi-dimensional parametric systems, different variables exhibit inherently different behaviors: parametric dependencies are often smooth and continuous; spatial modes reflect coherent structures; and temporal modes encode dynamical evolution with sequential correlations. A unified yet structured approximation framework that preserves multilinear organization while allowing mode-specific operator assignment remains largely unexplored.

In this work, we introduce MoTIF (Mode-structured Tensor Inference Framework), a structured tensor-based non-intrusive reduced-order modelling methodology designed for multi-parametric approximation, spatial resolution enhancement, and temporal forecasting within a unified formulation. The proposed framework is based on HOSVD to obtain a multilinear decomposition of high-dimensional datasets and adopts a mode-decoupled operator-learning strategy: Gaussian Process Regression is employed for parametric and spatial modal matrices, while recurrent neural networks are used to model temporal evolution.

The main contributions of this work are: (i) The formulation of a mode-structured tensor framework that preserves the intrinsic multilinear organization of multi-parametric datasets. (ii) The introduction of a decoupled operator-learning strategy that assigns dedicated approximation operators to each modal dimension. (iii) The unification of database completion, spatial super-resolution, and temporal forecasting within a single non-intrusive reduced-order modelling framework. (iv) The methodology is validated on a database of unsteady laminar flow simulations past a square cylinder under varying Reynolds numbers and angles of attack. Although demonstrated in a fluid dynamics context, the formulation is general and applicable to high-dimensional parametric systems governed by partial differential equations.

The remainder of this manuscript is organized as follows. Section 2 presents the proposed methodology and the structured tensor formulation. Section 3 describes the numerical database used for validation. Section 4 discusses the obtained results. Finally, Section 5 summarises the main conclusions of the work.

\section{Methodology} \label{Methodology}

This section provides a detailed description of the steps taken in the development of the fully data-driven ROM: MoTIF. These steps include the organization of the data obtained through CFD, an SVD-based temporal alignment method, and an in-depth explanation of the HOSVD modal decomposition technique. In addition, a detailed description is provided of the various machine learning approaches developed for data enhancement, the generation of databases under unseen flow conditions, and forecasting. Finally, the error metrics used to evaluate the proposed method against its counterpart are also presented.

\subsection{Conceptual overview of the proposed framework}

The proposed methodology is based on the structured representation of multi-dimensional parametric datasets as a tensor, and the subsequent assignment of dedicated approximation operators to each modal dimension, exploiting the multilinear structure characteristic of parametric dynamical systems.

The dimensions of this tensor correspond in order from left to right to: physical variables, parametric coordinates, spatial discretisation points, and time steps.HOSVD decomposes this tensor into a core tensor and a set of orthonormal mode matrices, one associated with each dimension.

Each mode matrix captures coherent structures associated with a specific dimension: physical parameters ( e.g. Reynolds number, Angle of Attack), spatial coordinates, and time. This framework adopts a mode-decoupled approximation strategy, where each modal dimension is treated independently and assigned a suitable operator depending on its mathematical and physical characteristics.

Parametric and spatial modes are approximated using Gaussian Process Regression, enabling smooth interpolation and controlled extrapolation in continuous domains. The temporal mode, which exhibits sequential dependencies and dynamical evolution, is modelled using recurrent neural networks. The velocity-component mode remains unchanged, preserving the physical interpretation of the field variables.

\begin{figure}
\centering
\includegraphics[width=0.75\linewidth]{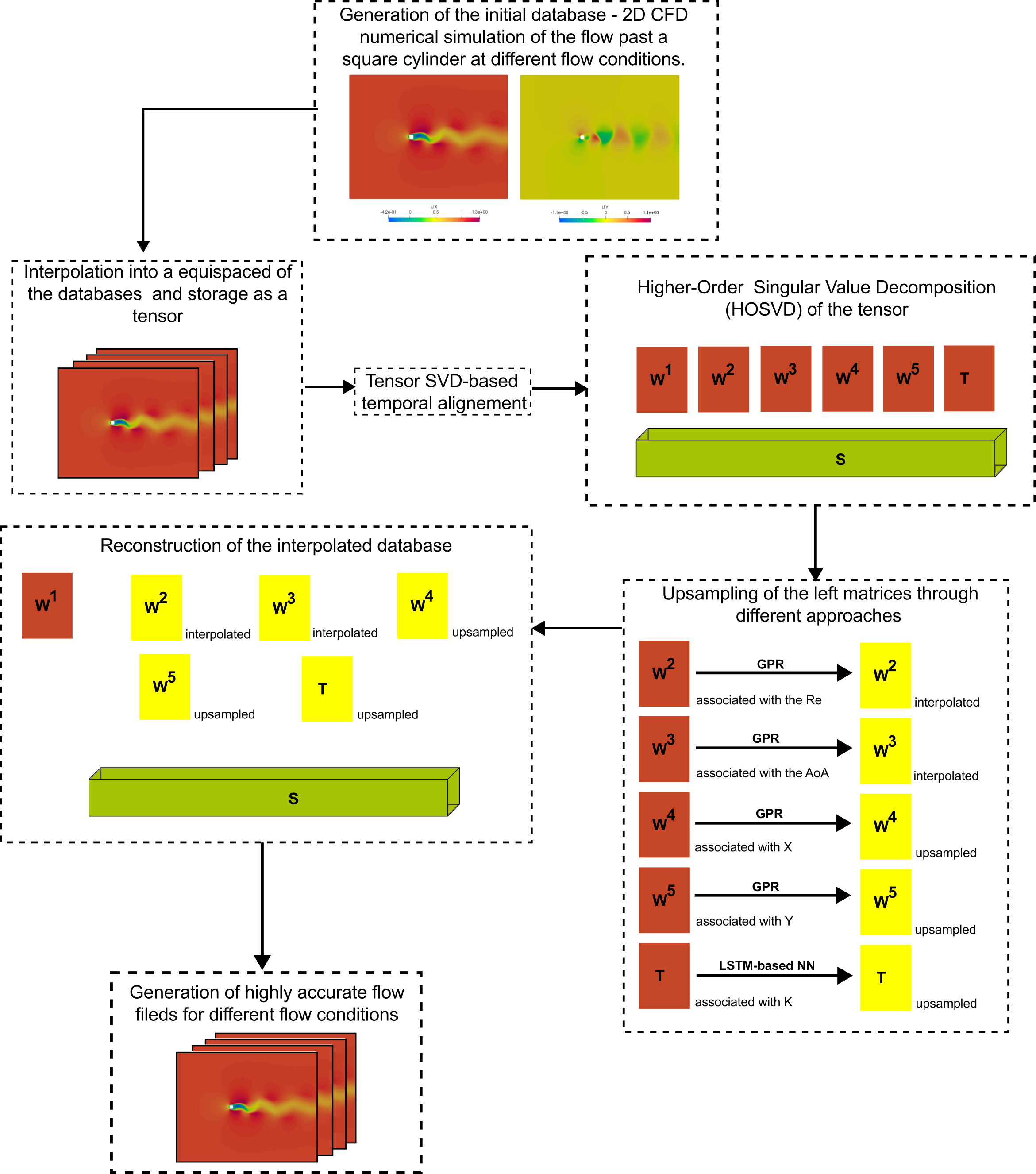}
\caption{ MoTIF framework.}
\label{overview_1}
\end{figure}

This operator-per-mode strategy provides several advantages, such as: (i)Preservation of the intrinsic multilinear tensor structure. (ii) Reduced regression complexity compared to monolithic surrogate models. (iii) Modular extensibility to additional parametric dimensions. (iv) Flexibility to incorporate alternative approximation operators.

The resulting framework constitutes a structured, non-intrusive reduced-order modelling approach capable of database completion, resolution enhancement, and temporal forecasting within a unified formulation.

\subsection{Data organization}

The fluid dynamics databases used in this work are organized as a multidimensional array, referred to as a \emph{snapshot tensor}, where the data are distributed on an equispaced two-dimensional grid. The flow-related information, such as the streamwise and normal velocity components, Reynolds number, angle of attack, spatial coordinates, and time, is stored separately within the tensor. This snapshot tensor is a collection of matrices, where the columns and rows for a given time correspond to the tensor \emph{fibres}. In this work, the two-dimensional flow past a square cylinder was first organized into fourth-order tensors and subsequently into sixth-order tensors.

For a two-dimensional time-dependent simulation defined by a particular Reynolds number and angle of attack, the streamwise and normal velocity components are represented in a \(J_2 \times J_3\) coordinate system (with $J_2$ and $J_3$ being the number of spatial grid points associated with each component, and $K$ the number of snapshots varying in time) as:

\begin{equation}
    \boldsymbol{v}(x_{j_2}, y_{j_3}, t_k) \quad \text{for } j_2 = 1, \dots, J_2, \quad j_3 = 1, \dots, J_3,~\text{and} \quad k = 1  \dots, K.
\label{4dimten}
\end{equation}

This information is organized as a fourth order \(J_1 \times J_2 \times J_3 \times K\)-tensor \(\mathbf{V}\), whose components \(V_{j_1 j_2 j_3 k}\) are defined as:

\begin{equation}
\label{equation:4dimtensor}
    \mathbf{V}_{1 j_2 j_3 k} = v_x (x_{j_2}, y_{j_3}, t_k), \quad 
    \mathbf{V}_{2 j_2 j_3 k} = v_y (x_{j_2}, y_{j_3}, t_k).
\end{equation}

where the index \( {j_1} \) refers to the streamwise  and normal  velocity components (\( {j_1} = 1,2\) where \(J_1 = 2\)). 
For cases where the database contains information under different flow conditions (such as varying Reynolds numbers and angles of attack, Re and AoA, respectively), the data must be organized as a sixth-order tensor ($J_1 \times Z_1 \times Z_2 \times J_2 \times J_3 \times K$). The flow condition data (\(\text{Re}, \text{AoA}\)) are contained in the \(Z_1\) and \(Z_2\) dimensions, respectively. The streamwise and normal velocity components are organized in a \(J_2 \times J_3\) coordinate system, as follows:

\begin{equation}
\begin{aligned}
\label{equation:6dimtensor}
    \mathbf{V}_{1 z_1 z_2 j_2 j_3 k} &= v_1 (Re_{z_1}, AoA_{z_2}, x_{j_2}, y_{j_3}, t_k), \quad
    \mathbf{V}_{2 z_1 z_2 j_2 j_3 k}   &= v_2 (Re_{z_1}, AoA_{z_2}, x_{j_2}, y_{j_3}, t_k), \\
\end{aligned}
\end{equation}

where the indexes \( {z_1} \) and \( {z_2} \) are related to the varying parameters \(Re\) and \(AoA\) that describe the flows. On the other hand, \({j_2} \) and \({j_3} \)  are the discrete values of the \(x\)- and \(y\)-axis, while \( k \) is the discrete time.

\subsection{POD Phase-Based Alignment}\label{Alignement}
In order to achieve consistency across simulations with varying unsteady behaviour (e.g., due to different Reynolds numbers and/or angles of attack), a phase-based alignment technique has been developed using SVD. This method maps the temporal evolution of each simulation onto a common, phase-aligned temporal grid based on the dominant flow dynamics.

Given a four-dimensional tensor, as in eq.~\eqref{4dimten}, where the first three dimensions correspond to the flow variables and the number of spatial points in the $x$ and $y$ directions, the data are reshaped into a two-dimensional matrix:

\begin{equation}
\mathbf{X} \in \mathbb{R}^{(J_1 J_2 J_3) \times K},   
\end{equation}
where $K$ denotes the number of temporal snapshots. The temporal mean is subtracted to obtain the fluctuation matrix:
\begin{equation}
\mathbf{X}' = \mathbf{X} - \bar{\mathbf{X}},
\end{equation}
where $\bar{\mathbf{X}}$ is the temporal mean computed along the second dimension.

Next, we perform SVD on the mean-subtracted matrix:
\begin{equation}
\mathbf{X}' = \mathbf{U} \mathbf{\Sigma} \mathbf{V}^\top,
\end{equation}
where $\mathbf{U} \in \mathbb{R}^{(J_1 J_2 J_3) \times r}$ contains the $r$ spatial SVD modes in its columns, $\mathbf{\Sigma} \in \mathbb{R}^{r \times r}$ is the diagonal matrix of singular values, and $\mathbf{V} \in \mathbb{R}^{K \times r}$ contains the temporal SVD coefficients, each one associated with a POD mode. Here, $r$ denotes the number of retained SVD modes (typically $r \ll \min(J_1 J_2 J_3, K)$).

A key property of SVD is that its spatial modes, the columns of $\mathbf{U}$, are ordered by decreasing energy, where the energy of the $i$-th mode is proportional to the $i$-th singular value $\sigma_{i}$. Hence, if $\mathbf{u}_i$ and $\mathbf{u}_j$ are the $i$-th and $j$-th columns of $\mathbf{U}$ (with corresponding singular values $\sigma_{i}$ and $\sigma_{j}$), then for any $i > j$,

\begin{equation}
    \sigma_{i} \geq \sigma_{j}.
\end{equation}
Therefore, the $i$-th SVD mode contains at least as much energy as the $j$-th mode.

\subsubsection{Phase alignment of the databases}

The first column of $\mathbf{V}$, denoted as $\mathbf{v}_1$, usually captures the dominant oscillatory behaviour in periodic or quasi-periodic flows, since it is related to the first (most energetic) SVD mode. This mode is normalized to lie within the interval $[-1, 1]$ using the following transformation:
\begin{equation}
\tilde{\mathbf{v}}_1 = 2 \cdot \frac{\mathbf{v}_1 - \min(\mathbf{v}_1)}{\max(\mathbf{v}_1) - \min(\mathbf{v}_1)} - 1.
\end{equation}

The normalized temporal coefficient $\tilde{\mathbf{v}}_1$ is used to determine the dominant period $T$ of the signal. This period is then used to phase-align all the databases, such that the onset of the dominant frequency cycle in the first SVD mode occurs at the same time instant across all databases. This alignment is crucial to ensure that the ROM can accurately predict the temporal dynamics of the system, as it removes phase discrepancies. For a given number of snapshots $n_{\text{snaps}}$, each period is discretized into a uniform temporal grid of K snapshots:
\begin{equation}
\Phi = \left\{ \phi_i = \frac{2\pi (i - 1)}{K}, \quad i = 1, \dots, K \right\}.
\end{equation}

The aligned temporal snapshots are obtained by extracting or interpolating the original temporal coefficients onto a structured phase grid. Cubic interpolation is applied when the target phase positions do not coincide with the original sampling points. This procedure ensures that all signal oscillations begin at the same phase and contain the same number of samples per cycle, which is essential for consistent phase alignment across databases.

\begin{figure}
\centering
\includegraphics[width=0.8\linewidth]{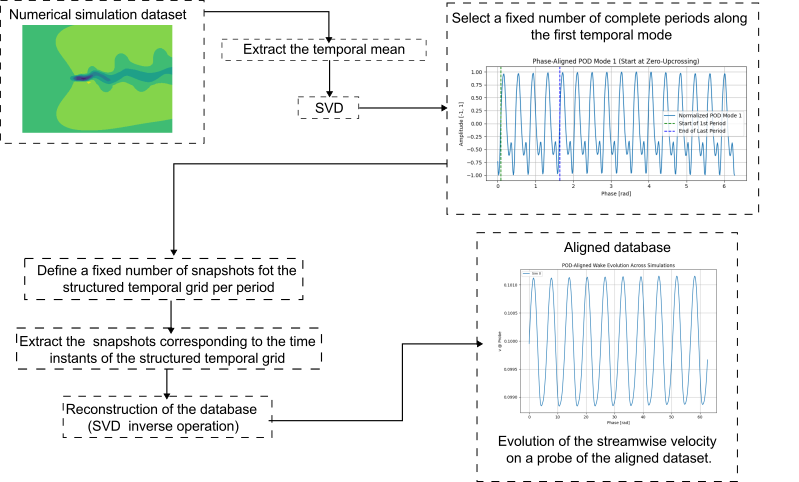}
\caption{ Sketch of the methodology applied for the SVD-based alignment of the databases obtained through numerical simulations.}
\label{fig: alignement}
\end{figure}

Finally, the aligned snapshot matrix $X_{\text{aligned}}$ is reconstructed using the aligned $r$ POD modes and reshaped back into its original four-dimensional spatial format with temporally aligned snapshots across the phase dimension, as shown in Fig. \ref{fig: alignement}

\subsection{Data pre processing}\label{preprocessing}
Before the dimensionality  step, normalization techniques must be applied to prevent the loss of relevant information associated with small magnitudes that may exist in the database. Centering and scaling normalization has been utilized in this work. Sola et al.~\cite{Sola1997} showed that adequate pre-processing of the data can reduce both the absolute error and the computational time during training by almost an order of magnitude~\cite{Peshwa} in deep learning architectures.

The combination of both centering and scaling transformations is also called standardization. This method ensures that the normalized output $\mathbf{V}_{\text{standardized}}$ has a mean equal to 0 and a standard deviation of 1~\cite{Huang2020}.

\begin{equation}
    \mathbf{V}_{\text{standardized}} = \frac{\mathbf{V} - \mu_\mathbf{V}}{\sigma_\mathbf{V}},
\end{equation}

where \(\mathbf{V}_{\text{standardized}}\) is the database, $\mu_{\mathbf{V}}$ is the mean of the database and $\sigma_{\mathbf{V}}$ the standard deviation. 

The centering and scaling pre-processing technique has been tested on all the databases included in this study. Centering and scaling were performed along the axes corresponding to variables, features, and time dimensions.

\subsection{Higher-order singular value decomposition}\label{HOVSD}
The High-order singular value decomposition (HOSVD) is an extension of SVD for multidimensional arrays. It was first introduced by Tucker in 1963 \cite{Tuck1963a,Tucker1966,Tuck2006} and later popularized around the year 2000 by De Lathauwer et al. \cite{DeLathauwer2000a,DeLathauwer2000b}. HOSVD has been applied in several engineering fields, such as aerodynamics~\cite{LORENTEcomp,Lorente2009HOSVD}, fluid dynamics~\cite{HETHERINGTON2024109217}, and reduced order modelling \cite{Moayyedi04072018}, among other things. For a six-dimensional snapshot tensor, the HOSVD can be expressed as:

\begin{equation}
\mathbf{V}_{j_1 z_1 z_2 j_2 j_3 k} \approx \sum_{p_1=1}^{P_1} \sum_{p_2=1}^{P_2} \sum_{p_3=1}^{P_3} \sum_{p_4=1}^{P_4} \sum_{p_5=1}^{P_5} \sum_{n=1}^{N} \mathbf{S}_{p_1 p_2 p_3 p_4 p_5 n} \mathbf{W}^{(1)}_{j_1 p_1} \mathbf{W}^{(2)}_{z_1 p_2} \mathbf{W}^{(3)}_{z_2 p_3} \mathbf{W}^{(4)}_{z_2 p_4} \mathbf{W}^{(5)}_{z_3 p_5} \mathbf{T}_{kn},
\label{equation: HOSVD}
\end{equation}

where $\mathbf{S}_{p_1 p_2 p_3 p_4 p_5 n}$ is referred to as the core tensor, $P_i$ refers to the number of retained singular vectors $i-th$ mode, and $\mathbf{W}^{(1)}$, $\mathbf{W}^{(2)}$, $\mathbf{W}^{(3)}$, $\mathbf{W}^{(4)}$, $\mathbf{W}^{(4)}$, and \(\boldsymbol{T}\) are the mode matrices resulting from the decomposition. $\mathbf{W}^{(1)}$ corresponds to the number of velocity components (i.e., streamwise and normal velocity components), while $\mathbf{W}^{(2)}$ and $\mathbf{W}^{(3)}$ represent the Re and AoA, respectively. $\mathbf{W}^{(4)}$ and $\mathbf{W}^{(5)}$ correspond to the mode matrices for the \(x\) and \(y\) spatial dimensions, respectively. The last matrix, $\mathbf{T}$, is associated with the temporal component. 

The mode matrices are obtained by applying SVD to each of the \emph{fibers} of the snapshot tensor. The set of SVD modes for each dimension corresponds to the eigenvectors associated with the positive eigenvalues. These SVD modes are orthonormal and are organized in decreasing order based on their energy content. The singular values are denoted as follows: 

\begin{equation}
\boldsymbol{\sigma}^{(1)}_{p_1}, \boldsymbol{\sigma}^{(2)}_{p_2}, \boldsymbol{\sigma}^{(3)}_{p_3}, \boldsymbol{\sigma}^{(4)}_{p_4}, \boldsymbol{\sigma}^{(5)}_{p_5}, \boldsymbol{\sigma}^{t}_{n}.
\label{HOSVD_modes}
\end{equation}

As mentioned above, HOSVD can be used for noise filtering and compression by truncating the SVD modes in each of the mode matrices obtained from the decomposition. This approach aims to extract the coherent structures of the flow, calculating the most relevant SVD modes associated with each component of the database. This results in a compressed approximation of the database. This approximation can be expressed as:

\begin{equation}
\mathbf{V}_{j_1 z_1 z_2 j_2 j_3 k} \approx \sum_{n=1}^{N} \mathbf{W}_{j_1 z_1 z_2 j_2 j_3 n} \hat{\mathbf{V}}_{kn},
\end{equation}

where N refers to the spatial complexity of the tensor which determines the number of spatial SVD modes  $\mathbf{W}_{j_1 z_1 z_2 j_2 j_3 n}$ and the rescaled temporal SVD modes \(\mathbf{\hat{V}}_{kn}\). The spatial SVD modes are defined as follows:  

\begin{equation}
\mathbf{W}_{j_1 z_1 z_2 j_2 j_3 n} = \sum_{p_1=1}^{P_1} \sum_{p_2=1}^{P_2} \sum_{p_3=1}^{P_3} \sum_{p_4=1}^{P_4} \sum_{p_5=1}^{P_5}  \mathbf{S}_{p_1 p_2 p_3 p_4 p_5 n} \mathbf{W}^{(1)}_{j_1 p_1} \mathbf{W}^{(2)}_{z_1 p_2} \mathbf{W}^{(3)}_{z_2 p_3} \mathbf{W}^{(4)}_{j_2 p_4} \mathbf{W}^{(5)}_{j_3 p_5}/{\mathbf{\sigma}^{t}_{n}}, 
\end{equation}

\begin{equation}
\hat{\mathbf{T}}_{kn} = \mathbf{\sigma}^{t}_{n} \mathbf{T}_{kn}.
\end{equation}

HOSVD is applied to the original six-dimensional tensor, retaining an appropriate, tunable number of SVD modes along each dimension. The number of retained modes is chosen to achieve an optimal balance between data compression, accurate reconstruction, and effective noise filtering. The resulting mode matrices for the Re (i.e., $\mathbf{W}^{(2)}$), the AoA (i.e., $\mathbf{W}^{(3)}$), the spatial dimensions (i.e., $\mathbf{W}^{(4)}$, $\mathbf{W}^{(5)}$), and the time coefficients $\mathbf{T}$, are therefore coupled to different machine learning approaches to expand the database in terms of flow parameters, space and time. The mode matrix for the velocity components (i.e., $\mathbf{W}^{(1)}$) remains unchanged. These approaches are detailed in the following section. 

The cumulative energy of the first $r$ singular values was calculated as a measure of how much of the total variance is captured by their corresponding modes. It is defined as:

\begin{equation}
E(r) = \frac{\sum_{i=1}^{r} \sigma_i^2}{\sum_{i=1}^{n} \sigma_i^2},
\label{eq:cum_energy}
\end{equation}

\noindent
where \( \sigma_i \) are the singular values of the data matrix, ordered such that \( \sigma_1 \geq \sigma_2 \geq \dots \geq \sigma_n \). The value \( E(r) \in [0, 1] \) quantifies the fraction of total variance (or energy) retained by the first \( r \) modes. This formulation is widely employed to identify the most energetic modes in a fluid dynamics databases~\cite{taira2017modal,holmes2012turbulence}.

\subsection{Gaussian Process Regression}

Gaussian Processes (GPs) are non-parametric supervised learning methods used for regression and probabilistic classification problems. Kolmogorov laid the foundation for GPs in 1938~\cite{kolmogorov}. Later, in 1978, Ibragimov \& Rozanov formally introduced them in Ref. \cite{Ibragimov}. In 2006, Rasmussen and Williams compiled previous theoretical developments on GPs in Ref. \cite{Rasmussen2006Gaussian} and applied them to regression and classification tasks, demonstrating the utility of Gaussian Process Regression (GPR) for both prediction and uncertainty quantification. GPR models the underlying function as a GP, allowing interpolation.

The prediction of a solution in an unexplored region for a database \( \mathbf{X} \) with \( N \) training points and corresponding outputs \( \mathbf{y} \) can be conducted using a GPR model. In general, the regression problem can be formulated as:

\begin{equation}
    y_i = f(\mathbf{x}_i) + \varepsilon, \quad \varepsilon \sim \mathcal{N}(0, \sigma_n^2),
\end{equation}

 \noindent{where \( \varepsilon \) represents Gaussian noise with variance \( \sigma_n^2 \) and mean $0$. Under the GP assumption, the function values follow a normal distribution during training, the function values follow a multivariate normal distribution with zero mean and covariance \( \mathbf{K} \):}

\begin{equation}
    \mathbf{f} \sim \mathcal{N}(\mathbf{0}, \mathbf{K}),
\end{equation}

 \noindent{where \( \mathbf{K} \) is the covariance matrix, defined as \( K_{ij} = k(\mathbf{x}_i, \mathbf{x}_j) \), representing the pairwise covariances between training inputs. The function \( f(\mathbf{x}) \) is modelled as a sample from a GP, which is defined as:}

\begin{equation}
    f(\mathbf{x}) \sim \mathcal{GP} \left( m(\mathbf{x}), k(\mathbf{x}, \mathbf{x}') \right),
\end{equation}

 \noindent{where \( m(\mathbf{x}) \) is the mean function, which is typically assumed to be \( 0 \), and \( k(\mathbf{x}, \mathbf{x}') \) is the covariance function (also referred to as the kernel), which determines the similarity between function values at different inputs.}

For computation with a test input \( \mathbf{x}_* \), the distribution of the observed outputs \( \mathbf{y} \) and the function value at \( \mathbf{x}_* \) is expressed as:

\begin{equation}
    \begin{bmatrix}
        \mathbf{y} \\
        f_*
    \end{bmatrix}
    \sim
    \mathcal{N} \left(
    \mathbf{0},
    \begin{bmatrix}
        \mathbf{K} + \sigma_n^2 \mathbf{I} & \mathbf{k}_* \\
        \mathbf{k}_*^\top & k(\mathbf{x}_*, \mathbf{x}_*)
    \end{bmatrix}
    \right),
\end{equation}

where \( \mathbf{k}_* = k(\mathbf{X}, \mathbf{x}_*) \) is the covariance vector between the training and test input values. The posterior predictive distribution, after applying standard Gaussian conditioning formulas, follows:

\begin{equation}
    p(f_* | \mathbf{X}, \mathbf{x}_*, \mathbf{y}) = \mathcal{N}(\mu_*, \sigma^2_*),
\end{equation}

 \noindent{where the mean prediction is given by:}

\begin{equation}
    \mu_* = \mathbf{k}_*^\top (\mathbf{K} + \sigma_n^2 \mathbf{I})^{-1} \mathbf{y},
\end{equation}

 \noindent{and the variance (uncertainty associated with the prediction) is expressed as:}

\begin{equation}
    \sigma^2_* = k(\mathbf{x}_*, \mathbf{x}_*) - \mathbf{k}_*^\top (\mathbf{K} + \sigma_n^2 \mathbf{I})^{-1} \mathbf{k}_*.
\end{equation}

 \noindent{The set of optimal hyperparameters for the Gaussian Process is defined as:}

\begin{equation}
    \theta = \{\sigma_f, \ell, \sigma_n\},
\end{equation}

 \noindent{where \( \sigma_f \) is the signal variance, which controls the amplitude of the function variations; \( \ell \) is the characteristic length scale, determining how quickly correlations decay with distance between input points; and \( \sigma_n \) is the standard deviation of the Gaussian noise associated with the observations. These hyperparameters are estimated by maximizing the marginal likelihood, expressed as:}

\begin{equation}
    \log p(\mathbf{y} | \mathbf{X}) = -\frac{1}{2} \mathbf{y}^\top (\mathbf{K} + \sigma_n^2 \mathbf{I})^{-1} \mathbf{y} - \frac{1}{2} \log |\mathbf{K} + \sigma_n^2 \mathbf{I}| - \frac{N}{2} \log 2\pi.
\end{equation}

\subsection{Neural networks based on LSTM architectures }

Neural Networks based on Long Short-Term Memory (LSTM)~\cite{LSTM} architectures are widely used tools for modelling and forecasting time series data~\cite{LINDEMANN2021650} and are particularly useful for data with complex temporal dynamics~\cite{HAJISHARIFI2024106361}. Generally, neural networks struggle with time series due to their inability to retain information over long sequences, LSTMs address this limitation through specialized memory cells and gating mechanisms namely, input, output, and forget gates which enable the network to selectively remember or discard information across time ~\cite{LSTM}. This allows LSTM networks to learn intricate temporal patterns and dependencies, making them especially effective for problems where the future state depends not only on recent inputs but also on long-term historical context. LSTMs can learn the underlying dynamics directly from data without requiring explicit knowledge of the governing equations.

\subsection{Development of the hybrid deep learning multi-parametric ROM: MoTIF}

\begin{figure}
\centering
\includegraphics[width=1\linewidth]{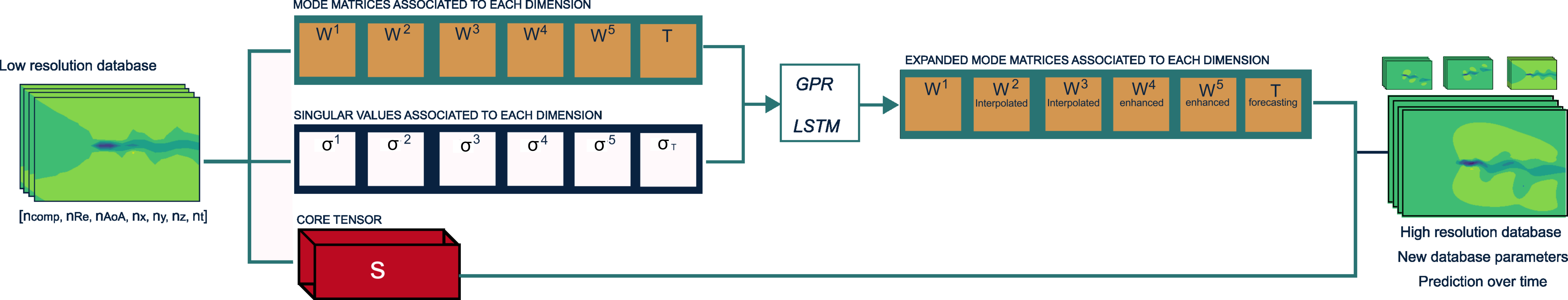}
\caption{\label{fig:methodology} Sketch of the methodology applied for the development of the proposed \textit{MoTIF}. Here, \textit{ncomp} denotes the number of velocity components, \textit{nRe} and \textit{nAoA} represent the number of Reynolds numbers and angles of attack in the database, respectively. \textit{nx} and \textit{ny} refer to the number of points along each spatial dimension, and \textit{t} indicates the number of temporal snapshots. \( W^{(i)} \) are the mode matrices associated with each dimension of the tensor, \( \sigma \) are the singular values, and \( \mathcal{S} \) denotes the core tensor.}
\end{figure}

This section outlines the methodology developed to construct a hybrid ROM capable of generating new flow condition databases, enhancing spatial resolution, and performing temporal forecasting (as shown in Fig.~\ref{fig:methodology}). The section is structured as follows: first, a general overview of the workflow is presented; next, the generation of new flow conditions using GPR is detailed; then, spatial resolution enhancement using ML and DL techniques is discussed; and finally, the temporal forecasting through DL models is addressed.

\subsubsection{Overview}
The development of the proposed MoTIF focuses on the offline stage, which involves all the above-mentioned pre-processing steps ~\ref{Alignement}, ~\ref{preprocessing}, HOSVD~\ref{HOVSD}, and its combination with different ML approaches. \newline{The initial step, the database is in tensor form, as in eq.~(\ref{equation:6dimtensor}), and it is decomposed using HOSVD.}

The resulting SVD mode matrices associated with the flow parameters, spatial, and temporal dimensions are used to expand the fluid dynamics multi-parametric database. MoTIF has been developed using a modular approach, where each module involves the use of a specific mode matrix (or set of mode matrices) with a specific ML approach and is independent of the others.

The database upscaling in terms of space is achieved by the application of GPR to the \( \mathbf{W^{(4)}} \) and \( \mathbf{W^{(5)}} \) spatial SVD mode matrices. The module developed for the generation of new flow conditions interpolates the SVD mode matrices \( \mathbf{W^{(2)}} \) and \( \mathbf{W^{(3)}} \), corresponding to the Re and AoA tensor components, using GPR. The forecasting module relies on an LSTM-based recurrent neural network applied to the temporal mode matrix, \( \mathbf{T} \).

The expansion of the spatial mode matrices, the interpolation over Reynolds number and AoA through GPR, and the training of the RNN are all conducted once during the initial stage (commonly referred to as the offline stage). In the subsequent stage (referred to as the online stage), the decomposed tensor, the expanded spatial mode matrices, the interpolated Reynolds number and AoA modes, along with the pre-trained neural network, are employed to perform predictions at low computational cost.

\subsubsection{Generation of  multi-parametric databases through GPR}

The generation of new databases containing information about fluid dynamics problems under different flow conditions, with varying parameters such as Re and AoA, is made possible through the combination of HOSVD and GPR interpolation. In this work, the GPR library \emph{scikit-learn}, available in \emph{Python}, has been used due to the robustness of its models. It is worth mentioning that other interpolation methods, such as linear interpolation, spline interpolation, and ordinary kriging, were tested during the development of the methodology, but GPR proved capable of capturing the non-linearities in the data, thereby improving the accuracy of the results.

\begin{figure}
\centering
\includegraphics[width=0.6\linewidth]{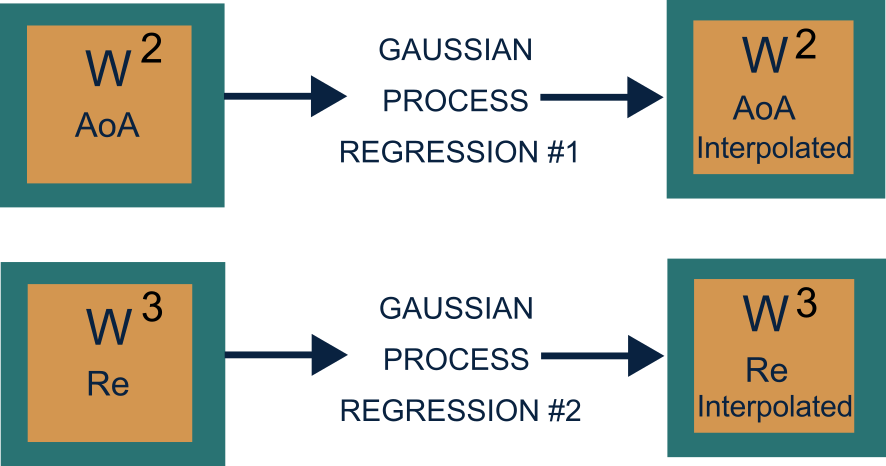}
\caption{ Sketch of the methodology applied for the generation of databases unseen flow conditions.}
\label{fig: GPR_met}
\end{figure}

The interpolation process follows the steps shown in Fig.~\ref{fig: GPR_met}. First, HOSVD is applied to the sixth-order tensor using Eq.~(\ref{equation: HOSVD}). After decomposition, the resulting mode matrices corresponding to the flow parameters Re and AoA, \( \mathbf{W}^{(2)} \) and \( \mathbf{W}^{(3)} \), are scaled by their associated singular values, \( \boldsymbol{\sigma}^{(2)} \) and \( \boldsymbol{\sigma}^{(3)} \), to preserve the contribution of each mode. These weighted mode matrices are then used as input for the GPR. The interpolation is performed independently for each available mode, or for each retained mode if truncation has been applied.

The GPR model uses several key parameters, which are listed in Tab.~\ref{tab:GPR_params}. The kernel defines the global scaling and determines how distances between data points are evaluated. The length scale, a component of the kernel, controls the smoothness of the predictions. The \( \alpha \) value is a small Gaussian noise term added to the diagonal of the covariance matrix to improve numerical stability and ensure proper conditioning during matrix inversion. Another important parameter is the number of optimizer restarts, which allows the model to explore different initial guesses and improve hyperparameter tuning.

After interpolation, the predicted values for each mode are combined with the original mode matrix. The full matrix is then reconstructed using the inverse HOSVD procedure.

\begin{table}
    \centering
    \begin{tabular}{llc}
        \toprule
        \textbf{Parameters} & & \textbf{Value} \\
        \midrule
        Kernel & Constant, RBF & 1 \\
        Length scale & & 50 \\
        $\alpha$ & &  \(1 \times 10^{-6}\) \\
        Restarts & & 5 \\
        \bottomrule
    \end{tabular}
    \caption{Summary of the GPR interpolation parameters.}
    \label{tab:GPR_params}
\end{table}

\subsubsection{Resolution enhancement through GPR}

\begin{figure}
\centering
\includegraphics[width=0.6\linewidth]{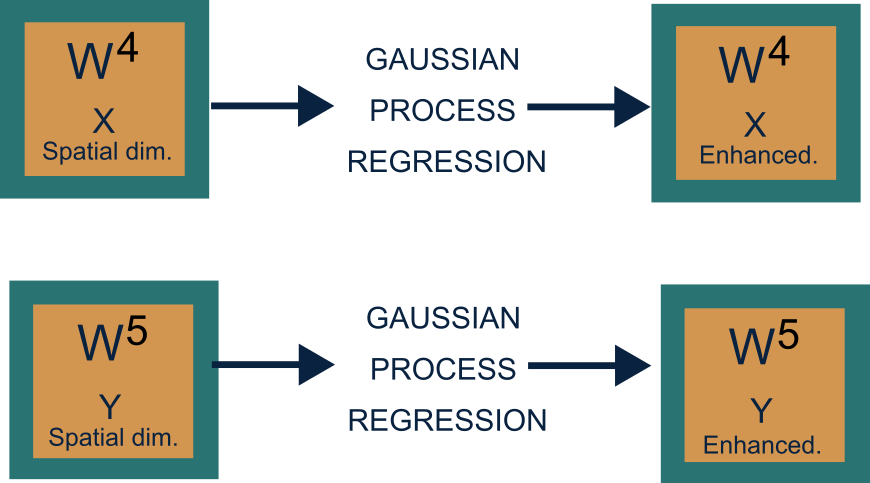}
\caption{\ Sketch of the methodology applied for the resolution enhancement using GPR.}
\label{fig:superresolution}
\end{figure}

As shown in Fig.~\ref{fig:superresolution}, the weighted spatial SVD mode matrices \(\mathbf{W}^{(4)}\) and \(\mathbf{W}^{(5)}\), resulting from the application of HOSVD to a low-resolution database with \emph{m} SVD modes and \emph{n} low-resolution spatial points (i.e., fewer points in the spatial dimensions), serve as the inputs for another GPR, the parameters of which are detailed in Tab.~\ref{tab:gprsr}. In this approach, a \emph{Matern} kernel, which is a generalization of the RBF kernel, has been used. This kernel specifies the covariance between two measurements as a function of their distance~\cite{Rasmussen2006Gaussian}, and the smoothness of the interpolation is controlled by the parameter \(\nu\); larger values of \(\nu\) lead to smoother functions.

\begin{table}
    \centering
    \begin{tabular}{llc}
        \toprule
        \textbf{Parameters} & Type & \textbf{Value} \\
        \midrule
        Kernel & Constant, Matern & 1 \\
        Length scale & & 0.05 \\
        $\alpha$ & &  \(1 \times 10^{-6}\) \\
        $\nu$ & &  \(1.5\) \\
        Restarts & & 20 \\
        \bottomrule
    \end{tabular}
    \caption{Summary of the GPR spatial resolution enhancement parameters}
    \label{tab:gprsr}
\end{table}

\subsubsection{Temporal forecasting through Deep Learning}

The prediction of fluid dynamics behavior has been achieved through the combination of HOSVD and deep neural networks. In this study, the \emph{Keras} API~\cite{chollet2015keras} is employed to develop a simple yet robust recurrent neural network, RNN. The proposed approach is autoregressive, meaning that the neural network's predictions are recursively used as inputs for future predictions. This enables long-term forecasting with reduced computational time. The generated data can then be used by MoTIF as a ROM, of the CFD simulation, thereby accelerating the numerical simulation process.

\begin{figure}
\centering
\includegraphics[width=1\linewidth]{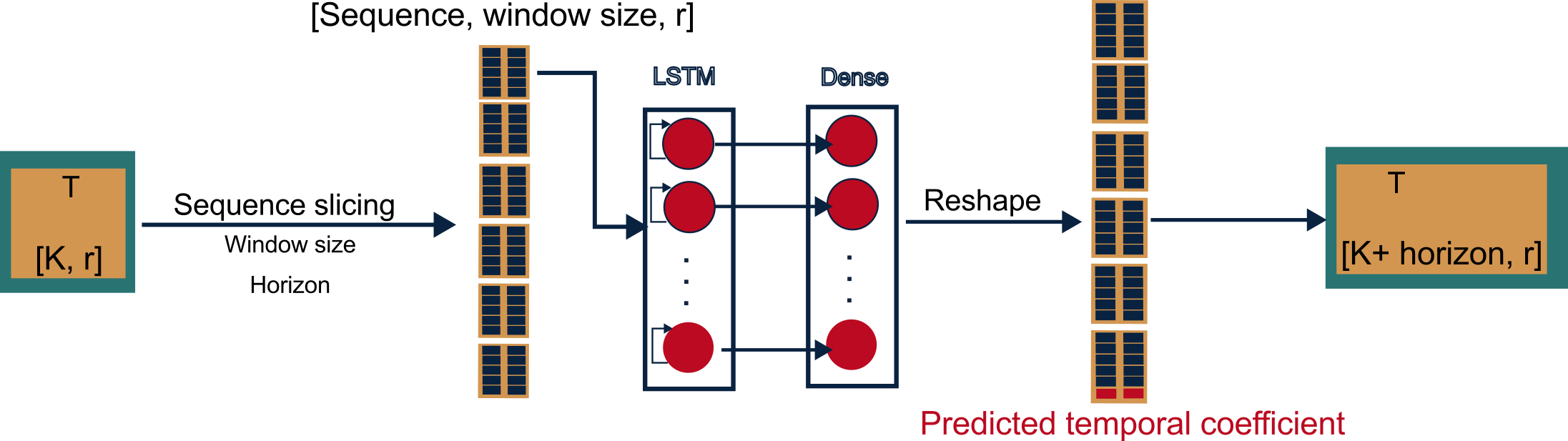}
\caption{\label{fig:forecasting} Sketch of the methodology used for forecasting.}
\end{figure}

Figure~\ref{fig:forecasting} illustrates the steps taken to forecast the behaviour of a fluid dynamics problem. First, the weighted SVD mode matrix associated with the temporal dimension and calculated through HOSVD, denoted as \(\mathbf{T}\), is extracted. Each SVD mode is then treated as a time series. To train the RNN, these time series are segmented into multiple input sequences. Each sequence has a fixed \emph{length} (i.e., the number of time values used as input), and the network is trained to predict a future time window defined by the prediction horizon \(h\). These input sequences serve as the training data for the neural network described in the following paragraph.

The RNN architecture developed in this work is simple yet robust. It consists of one LSTM layer~\cite{hochreiter1997long}, followed by a fully connected layer to map the output to the desired dimensions, and a reshape layer to ensure consistency in output structure. The Adam optimizer~\cite{2017_Kingma_etal_Adam} is used for training. Mean Squared Error (MSE) is employed as the loss function during training, while Mean Absolute Error (MAE) is used for validation. The number of LSTM units, neurons, and other details are summarized in Table~\ref{tab:forecasting}.

The LSTM layer was selected due to its ability to learn long-term dependencies in time series data. This is achieved through a gating mechanism that controls the flow of information. It includes three main gates: the forget gate, the input gate, and the output gate. The forget gate determines which past information should be discarded; the input gate decides which new information should be stored; and the output gate regulates the information passed to the next time step~\cite{hochreiter1997long}.

The temporal matrix \(\mathbf{T}\) is split into three subsets for training (70\% of the total data), validation (15\%), and testing (15\%), following standard machine learning practices.

\begin{table}
    \centering
    \begin{tabular}{llc}
        \toprule
        \textbf{Layer} & \textbf{Parameter} & \textbf{Value} \\
        \midrule
        LSTM & Units & 128 \\
        Dense & Neurons & 10 \\
        Reshape & Shape & (horizon = 1, 10) \\
        \midrule
        \multicolumn{2}{l}{\textbf{Hyperparameters}} & \\
        \midrule
        & Learning rate & 0.001 \\
        & Optimizer & Adam \\
        & Batch size & 8 \\
        \bottomrule
    \end{tabular}
    \caption{Summary of the hyperparameters utilised during the training of the RNN implemented in this study to perform the forecast of a fluid dynamics problem.}
    \label{tab:forecasting}
\end{table}

\subsubsection{Error metrics} \label{error}

IN this work, we have considered two error measures: 

\paragraph{Relative root mean squared error - RRMSE }

The Relative Root Mean Square Error (RRMSE) is a metric used to evaluate the accuracy of the proposed approaches by comparing the predicted tensor reconstruction \( \mathbf{\hat{V}} \) with the reference tensor \( \mathbf{V} \) obtained from CFD simulations. This comparison is performed for a specific Reynolds number, AoA, spatial resolution, and forecast horizon. The RRMSE is defined as:
\begin{equation}
\text{RRMSE} = \frac{\sqrt{\sum_{i=1}^{N} \left( \mathbf{V}_i - \mathbf{\hat{V}}_i \right)^2}}{\sqrt{\sum_{i=1}^{N} \mathbf{V}_i^2}},
\label{RRMSE_eqn}
\end{equation}
where \( \mathbf{V}_i \) and \( \mathbf{\hat{V}}_i \) represent the components of the ground truth and predicted tensors, respectively, and \( N \) is the number of snapshots. A lower RRMSE value indicates higher reconstruction accuracy.

\paragraph{Estimated probability density function of the normalized error}

The normalized error is defined as the difference between the ground truth tensor $\boldsymbol{V}$ and the reconstructed tensor $\boldsymbol{V}^{US}$. This value is then normalised between $0$ and $1$ by the maximum absolute error and can be expressed as:
\begin{equation}
\epsilon_{i} = \frac{\mathbf{V}_i - \mathbf{V}^{US}_i}{\max(|\mathbf{V}_i - \mathbf{V}^{US}_i|)},
\end{equation}
where \( \mathbf{V}_i \) is the real value, \( \mathbf{V}^{US}_i \) is the predicted value, and \( \max(|\mathbf{V}_i - \mathbf{V}^{US}_i|) \) is the maximum absolute error among all data points.

The distribution of the normalized error of the predicted database is displayed in a relative frequency histogram, where the optimal number of bins or classes for a histogram have been calculated using the Sturges rule, expressed as: 
\begin{equation}
    k_{bins} = 1 + log_2(N), 
\end{equation}
where N corresponds to the number of spatial points per snapshot. 

\section{Database}\label{model}

The validation of the methodology has been performed using a database composed of a set of two-dimensional numerical simulations that recreate the behaviour of a laminar flow passing over a square cylinder. This section details the governing equations of the problem under study, the set-up of the numerical simulation in regards of the computational domain, grid development, and boundary conditions. 

\begin{figure}
\centering
\includegraphics[width=0.35\linewidth]{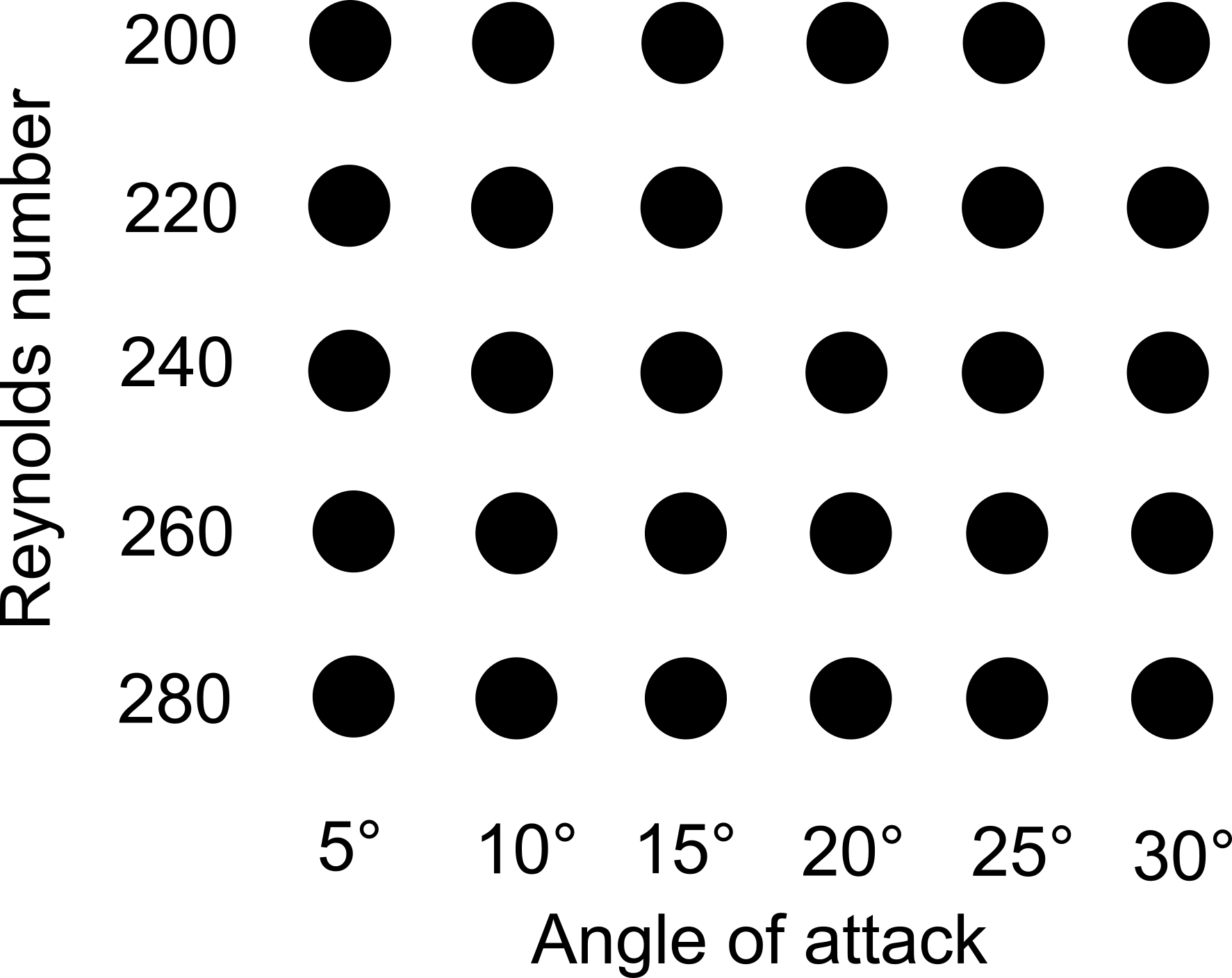}
\caption{\label{fig:designspace} Design space distribution, each black dot represents a database obtained through numerical simulations for a specific Re and AoA.}
\end{figure}

The results obtained through the numerical simulations of the two-dimensional laminar flow over a square cylinder are finally organized in a 6th-order tensor, as in eq. \eqref{equation:6dimtensor}. The characteristics of this tensor are detailed in Tab. \ref{tab: dimensions_tensors}. Figure \ref{fig:designspace} displays the design space of the database, the generated databases for unseen flow conditions must be inside these limits. 

\begin{table}
    \centering
    \begin{tabular}{ccccccc}
        \toprule
         Tensor order &  $n_{var}$ &  $n_{Re}$ &  $n_{AoA}$ & $n_x$ & $n_y$ & K \\
        \midrule
       
        6 & 2 & 6 & 6 & 64 & 64 & 128 \\
       
        \bottomrule
    \end{tabular}
    
    \caption{Characteristics of the resulting database. Here, \( n_{\text{var}} \) corresponds to the streamwise and normal velocity components, \( n_{\text{Re}} \) to the amount of Re numbers on the database, \( n_{\text{AoA}} \) to the amount of AoA on the database, \( n_x \) and \( n_y \) to the spatial points in the \( x \)- and \( y \)-dimensions, respectively, and \( K \) to the number of available snapshots. Refer to Fig. \ref{fig:designspace} for further details about the flow conditions.}
\label{tab: dimensions_tensors}
\end{table}

\begin{figure}
    \centering
    \begin{subfigure}[b]{0.35\textwidth}
        \centering
        \includegraphics[width=\textwidth]{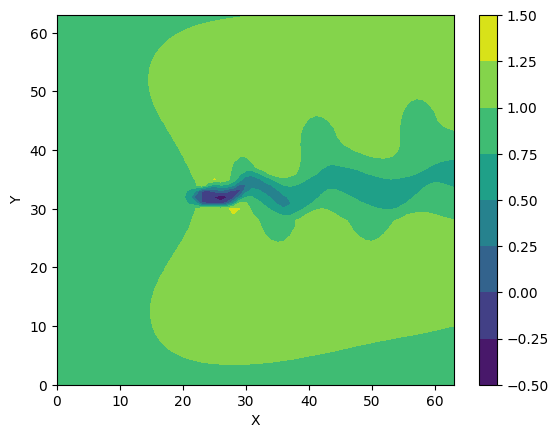}
        \caption{}
        \label{fig:5_200}
    \end{subfigure}
    \hspace{1cm}
    \begin{subfigure}[b]{0.35\textwidth}
        \centering
        \includegraphics[width=\textwidth]{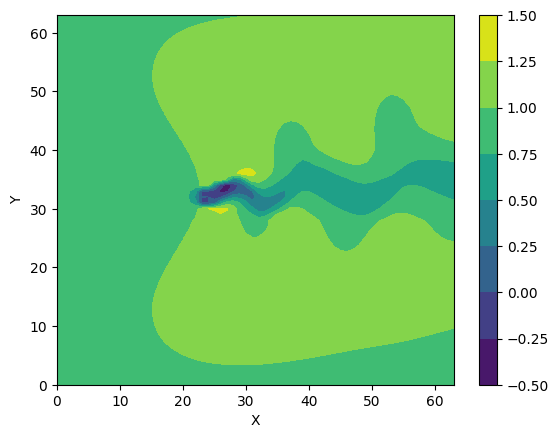}
        \caption{}
        \label{fig:5_280}
    \end{subfigure}
    \vfill
    \begin{subfigure}[b]{0.35\textwidth}
        \centering
        \includegraphics[width=\textwidth]{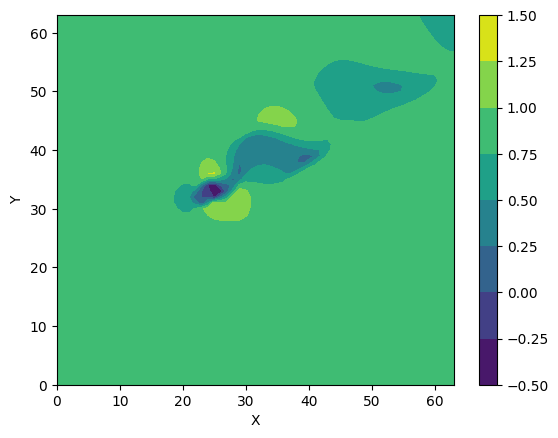}
        \caption{}
        \label{fig:30_200}
    \end{subfigure}
    \hspace{1cm}
    \begin{subfigure}[b]{0.35\textwidth}
        \centering
        \includegraphics[width=\textwidth]{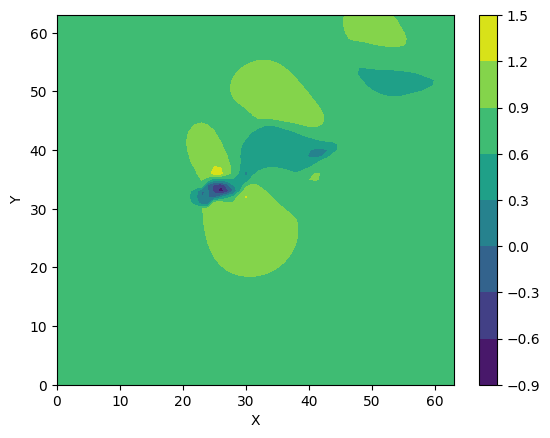}
        \caption{}
        \label{fig:30_280}
    \end{subfigure}
    \hfill

    \caption{Streamwise velocity field of the database at different flow conditions: (a) Re = 200 $\&$ AoA = $5^\circ$, (b) Re = 280 $\&$ AoA = $5^\circ$, (c) Re = 200 $\&$ AoA = $30^\circ$, (d) Re = 280 $\&$ AoA = $30^\circ$}
    \label{fig:data_stream}
\end{figure}

\subsection{Flow around a square cylinder }

The flow around square cylinders is a widely studied problem in fluid dynamics. It is characterized by flow separation at the sharp corners of the square cylinder, creating low-pressure zones behind the cylinder that promote the formation of vortex shedding. At Reynolds numbers above 4000, this vortex shedding becomes turbulent, leading to aleatory fluctuations in the flow patterns~\cite{TRIAS201587}.

The flow around square cylinders has been selected as a benchmark problem for this study due to the extensive research conducted on this phenomenon under various flow conditions over the past decades. This background enables the analysis of different flow parameters such as AoA and Re.  Variations in these parameters are known to induce significant changes in the flow topology, including alterations in the onset and location of flow separation, modifications in the wake structure, transition between steady and unsteady regimes, and changes in vortex shedding frequency and symmetry~\cite{BELTRAN2019105428}. Studying the flow around a square cylinder is particularly relevant as it serves as a simplified model for understanding flow behaviour around bluff bodies, similar to those encountered in urban environments.

\subsection{Governing equations}

The governing equations for an incompressible, two-dimensional Newtonian flow are given by the continuity and  Navier-Stokes equations, written as: 

\begin{equation}
\nabla \cdot \mathbf{v} = 0, \quad \frac{\partial \mathbf{v}}{\partial t} + (\mathbf{v} \cdot \nabla) \mathbf{v} = -\nabla p + \frac{1}{Re} \nabla^2 \mathbf{v}.
\end{equation}

where \( \mathbf{v} \) is the non-dimensional velocity vector containing the streamwise and normal components in a two-dimensional setting, defined as \( (v_x, v_y) \), and \( p \) is the pressure. Length, time, velocity, and pressure are non-dimensionalized using \( L_c^* \), \( L_c^*/V_c^* \), \( V_c^* \), and \( \rho^* V_c^{*2} \), respectively. The Reynolds number is given by \( \mathrm{Re} = V_c^* L_c^* / \nu^* \), where the superscript \( * \) indicates dimensional quantities. Here, \( L_c^* \) is the side length of the square, and \( V_c^* \) is the incoming free-stream velocity.

\subsection{Geometry, computational domain, mesh and boundary conditions}

\begin{figure}
\centering
\includegraphics[width=0.7\linewidth]{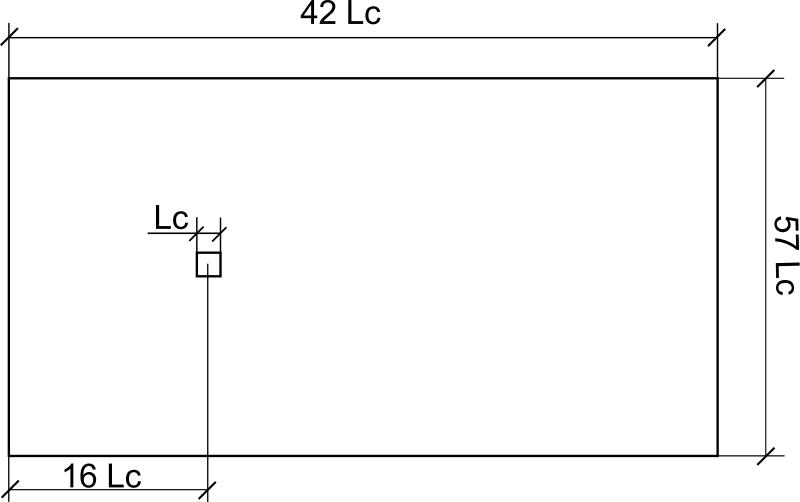}
\caption{\label{fig:computational_domain} Sketch of the computational domain dimensions for the two-dimensional numerical simulation of a laminar flow over a square cylinder.}
\end{figure}

For these two-dimensional simulations, the domain geometry has been developed in a Cartesian coordinate system in the X-Y plane, following the criteria of ~\cite{BELTRAN2019105428}. The characteristic length of the square cylinder is referred to as ``$L_c$" and was used to adimensionalize the domain. The distance considered for the upstream is $16L_c$ while for the downstream is $25L_c$. The distance between the superior and inferior boundaries is $57L_c$. Figure~\ref{fig:computational_domain}, illustrates the distribution in the computational domain. For the above-described configuration, a set of numerical simulations will be performed for flows with various Reynolds numbers in laminar regime (Re from 200 to 600 in steps of 20) and also for a different set of AoA (values from  $5{\circ}$ to $30^{\circ}$ in steps of $5^{\circ}$).

A structured grid consisting of 112,000 hexahedral elements was generated using Gmsh 4.13.1, following the meshing methodology proposed by \cite{10.1063/5.0025201}, which is recognized for its high accuracy in laminar flow simulations. This grid resolution is sufficient to construct a high-quality database for the evaluation of MoTIF. The resulting database supports three key capabilities: (i) enhancing spatial resolution, (ii) interpolating to infer flow fields at previously unseen Reynolds numbers and angles of attack, and (iii) producing time-accurate predictions. The primary objective of this work is to demonstrate the practical application of MoTIF, which is introduced here for the first time.

\begin{figure}
\centering
\includegraphics[width=0.75\linewidth]{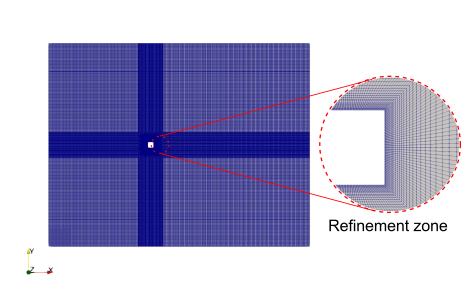}
\caption{\label{fig:mesh_refinements} Structured mesh developed for the two-dimensional numerical simulation of the laminar flow past a square cylinder.}
\end{figure}

\begin{figure}[H]
\centering
\includegraphics[width=0.65\linewidth]{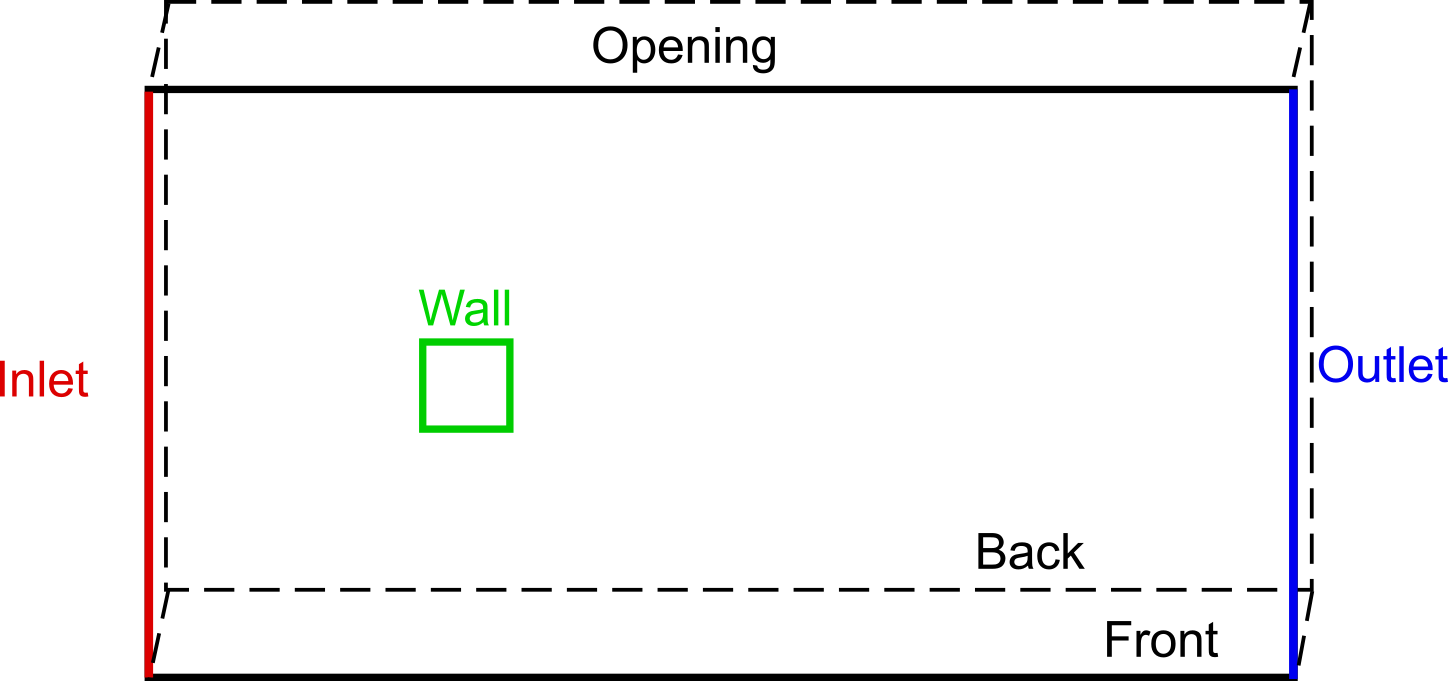}
\caption{\label{fig:boundary_conditions} Sketch of the computational domain dimensions and boundary conditions. A fixed velocity is prescribed at the inlet based on the angle of attack (AoA), with \( U_x = U_{\infty} \cos(\text{AoA}) \), \( U_y = U_{\infty} \sin(\text{AoA}) \), and \( U_{\infty} = 1 \). At the outlet, pressure is fixed to zero and a zero-gradient condition is applied to velocity. Top and bottom boundaries are treated as open-flow with zero-gradient velocity and fixed pressure.}

\end{figure}

The numerical simulations have been performed using OpenFOAM 9~\cite{JASAK200989}, an open-source CFD solver based on the finite volume method for spatial discretization. The temporal discretization employs a second-order Euler scheme, while the pressure-velocity coupling is handled using the PIMPLE algorithm, which is a combination of PISO~(Pressure Implicit with Splitting of Operator) and SIMPLE (Semi-Implicit Method for Pressure-Linked Equations) methods. The simulation has been performed using the PIMPLE algorithm,  which is used for transient simulations. The boundary conditions were set as follows: a Dirichlet condition was imposed for the velocity at the inlet, where the streamwise and crosswise components were defined as $U_x = U_\infty \cos(\text{AoA})$, $U_y = U_\infty \sin(\text{AoA})$, with $U_\infty = 1$. At the outlet, a Dirichlet condition for pressure ($p = 0$) and a zero-gradient (Neumann) condition for velocity were applied, allowing for outflow. On the top and bottom boundaries, zero-gradient conditions for velocity and fixed pressure were imposed as open-flow conditions. The Reynolds number was controlled by varying the dynamic viscosity of the fluid while maintaining a constant free-stream velocity. Variations in AoA were introduced by adjusting the inlet velocity vector. The time step has been established as $\Delta$ t = 0.001 [s], with a total simulation time of 5 [s] with a writing interval of the flow variables of 0.02 [s]. Figure ~\ref{fig:boundary_conditions} depicts the boundary conditions set in the computational domain.

\section{Results and discussion}\label{results}

This section presents the results obtained from applying the proposed MoTIF for resolution enhancement, forecasting, and multi-parametric database generation. This section is organized as follows: first, an analysis of the singular value decay for each of the SVD mode matrices associated with Re, AoA, and temporal dimensions is provided, together with the GPR interpolation results. Then, the results obtained for the four different test cases considered for the validation of this method are discussed. Finally, a summary of the prediction of RRMSE is presented.  

To demonstrate the efficiency, robustness, and accuracy of the proposed tool, two different scenarios have been considered: the completion of inconsistent multi-parametric databases and the generation of databases for unseen flow conditions. These results have been evaluated using the error metrics detailed in Sec.~\ref{error}. Table~\ref{tab: testcases} provides a detailed overview of the aforementioned test cases, the techniques employed, and their corresponding scenarios. Table \ref{tab:rrmse} collects the prediction RRMSE, calculated as in eq. \eqref{RRMSE_eqn} of the test cases detailed in Tab. \ref{tab: testcases}.

\begin{table}
    \centering
    \begin{tabular}{cccccc}
        \toprule
         ID & Task (interpolation) & Desired &  Desired & Res. enhance. & Forecasting   \\
          &  &  Re & AoA  & upscaling  &  snaps  \\
          &  &  &   & factor.  &   predicted.  \\
         
        \midrule
       
        FAoA & Filling missing data in AoA dim. & 240 &  $15^\circ$  & 2 & 100\\
        FRe & Filling missing data in Re dim. & 260 &  $20^\circ$ & 2 & 100 \\
        N1 & Generate data for unseen conditions & 230 &  $22.5^\circ$  & 2 & 100\\
        N2 & Generate data for unseen conditions & 245 &  $11^\circ$ & 2 & 100 \\
       
        \bottomrule
    \end{tabular}
    
    \caption{Overview of the test cases used to validate the proposed MoTIF. The ID column provides a short identifier for each case. The Task describes the interpolation or prediction objective (e.g., missing data reconstruction or generation of unseen conditions). The columns Desired Re and Desired AoA indicate the Reynolds number and angle of attack used as targets for reconstruction. The Res. enhance. upscaling factor specifies the temporal resolution enhancement factor applied to the data. The Forecasting snaps predicted denote the number of future snapshots predicted during the testing phase.}

    \label{tab: testcases}   
\end{table}

\subsection{HOSVD analysis and GPR interpolation for Re and AoA}

As mentioned in Section~\ref{HOVSD}, the singular values \(\sigma_i^{(1,2)}\) (Eq.~\eqref{HOSVD_modes}) associated with each dimension of the tensor result from applying HOSVD to the database. To perform dimensionality reduction and determine how many modes to retain in each mode matrix, the energy content of the singular values was analyzed along the Re dimension (\(\sigma^{(1)}\)) and the AoA dimension (\(\sigma^{(2)}\)).

\begin{figure}
    \centering
    \begin{subfigure}[b]{0.48\textwidth}
        \centering
        \includegraphics[width=\textwidth]{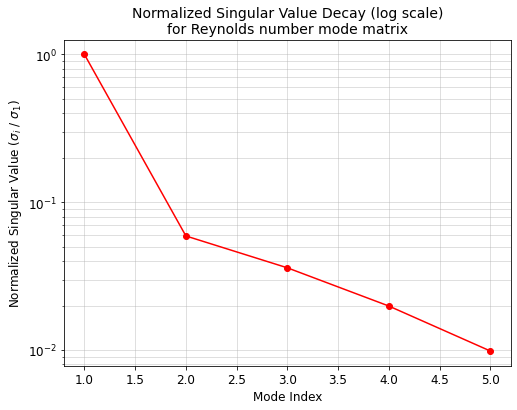}
        \caption{}
        \label{fig:Re_sv}
    \end{subfigure}
    \hfill
    \begin{subfigure}[b]{0.48\textwidth}
        \centering
        \includegraphics[width=\textwidth]{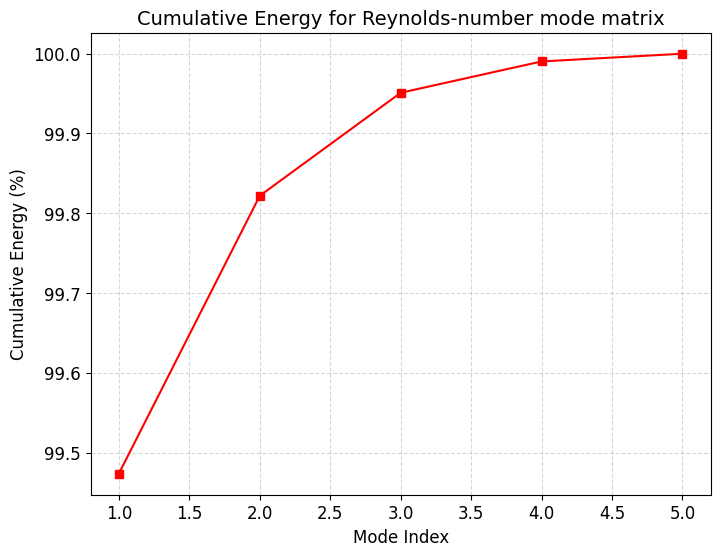}
        \caption{}
        \label{fig:Re_cum}
    \end{subfigure}
    \caption{Singular value decay (a), and cumulative energy (b) plots for the SVD mode matrix associated with the Re dimension.}
    \label{fig:Re_sv1}
\end{figure}

For the Re dimension, the associated singular values \(\sigma^{(1)}\), the cumulative energy content per SVD mode, and the singular value decay are displayed in Fig.~\ref{fig:Re_sv1}. Figure~\ref{fig:Re_sv} shows a significant drop in singular value magnitudes between the first and second modes, forming a sharp elbow. This suggests that the first two singular values are considerably more important than the rest. The same trend is seen in the cumulative energy plot (Fig.~\ref{fig:Re_cum}), where the first two singular values account for 99.8\% of the total energy. As a result, dimensionality reduction can be applied by keeping only the first three modes, which together capture more than 99.9\% of the total energy.

\begin{figure}
    \centering
    \begin{subfigure}[b]{0.48\textwidth}
        \centering
        \includegraphics[width=\textwidth]{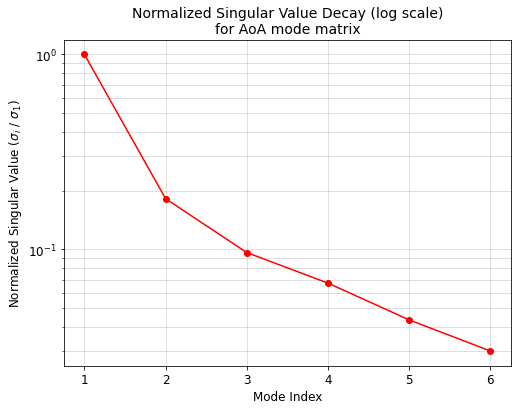}
        \caption{}
        \label{fig:Aoa_sv}
    \end{subfigure}
    \hfill
    \begin{subfigure}[b]{0.48\textwidth}
        \centering
        \includegraphics[width=\textwidth]{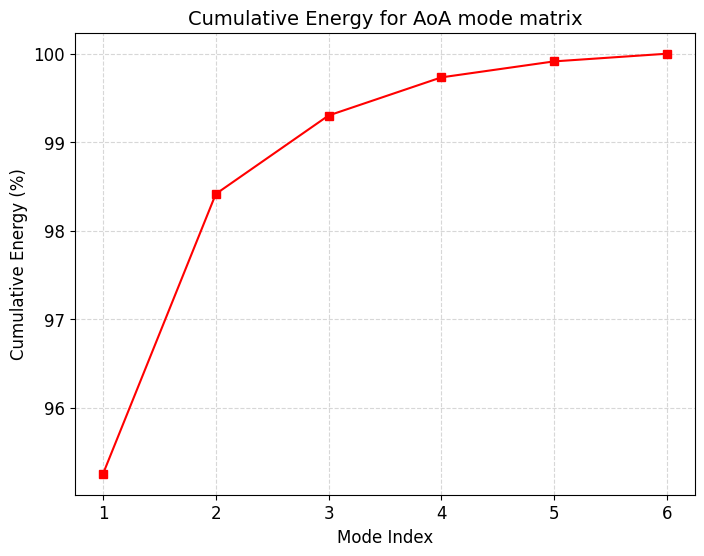}
        \caption{}
        \label{fig:Aoa_cum}
    \end{subfigure}
    \caption{Same as Fig. \ref{fig:Re_sv}, but for the AoA SVD mode matrix.}
    \label{fig:Aoa_sv0}
\end{figure}

Regarding the AoA dimension, Fig.~\ref{fig:Aoa_sv} shows a significant decay in the \(\sigma^{(2)}\) magnitudes between the first and second modes, followed by a more gradual decline in the subsequent modes. In contrast to the Re dimension (Fig.~\ref{fig:Re_sv1}), the cumulative energy plot in Fig.~\ref{fig:Aoa_cum} indicates that 96\% of the energy is captured by the first mode. However, to exceed 99\% of the total energy, at least four modes must be retained.
It is important to note that only 5 Reynolds numbers are available in the database. Therefore, the input matrix for the Re dimension has a rank of 5, which limits the maximum number of singular values to 5. To ensure accurate reconstruction, the first four modes along the AoA dimension have been selected, accounting for over 99.5\% of the total energy.

\begin{figure}
    \centering
    \begin{subfigure}[b]{0.48\textwidth}
        \centering
        \includegraphics[width=\textwidth]{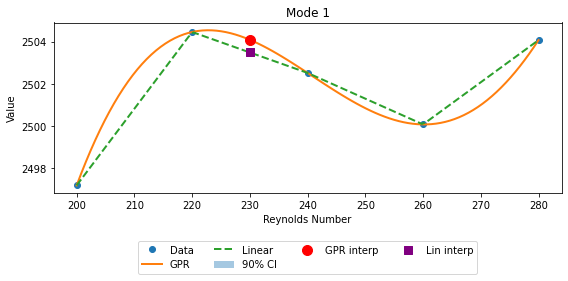}
        \caption{}
        \label{fig:Re_0}
    \end{subfigure}
    \hfill
    \begin{subfigure}[b]{0.48\textwidth}
        \centering
        \includegraphics[width=\textwidth]{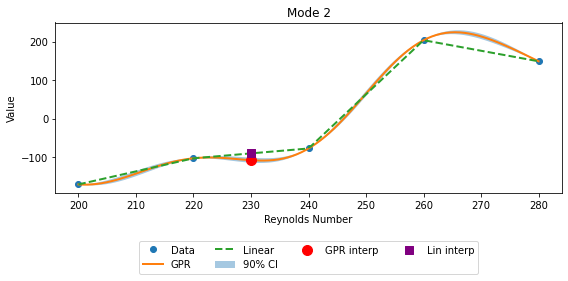}
        \caption{}
        \label{fig:Re_1}
    \end{subfigure}

    \vskip\baselineskip

    \begin{subfigure}[b]{0.48\textwidth}
        \centering
        \includegraphics[width=\textwidth]{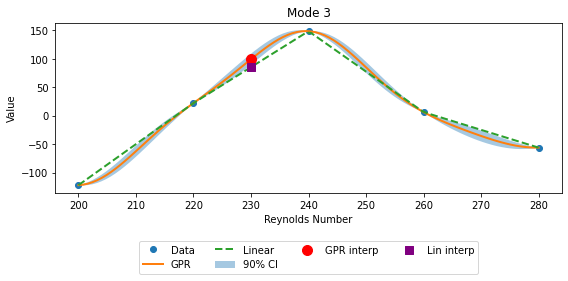}
        \caption{}
        \label{fig:Re_2}
    \end{subfigure}
    \hfill

    \caption{GPR and piecewise linear interpolation for the first 3 SVD modes calculated associated to the Re number dimension.}
    \label{fig:Re_gpr}
\end{figure}

Figure~\ref{fig:Re_gpr} compares the regression results obtained using Gaussian Process Regression (GPR) and linear interpolation across each of the retained modes of the SVD mode matrix associated with the Reynolds number dimension, \(\mathbf{W}^{(2)}\). GPR provides a smooth, non-linear fit to the data, with a narrow region representing the 90\% confidence interval, as shown in Fig.~\ref{fig:Re_1} and Fig.~\ref{fig:Re_2}. In contrast, the piecewise linear interpolation consists of straight segments that fail to capture the underlying curvature or non-linear trends, thereby missing key variations in the data. Multiple reconstruction tests were conducted using both methods, and GPR consistently produced more accurate results due to its ability to model non-linear behaviour. Notably, in Fig.~\ref{fig:Re_0}, the confidence interval region is barely visible. This results from the GPR model fitting the underlying function with high accuracy and low uncertainty, leading to a very narrow confidence interval.

\begin{figure}
    \centering
    \begin{subfigure}[b]{0.48\textwidth}
        \centering
        \includegraphics[width=\textwidth]{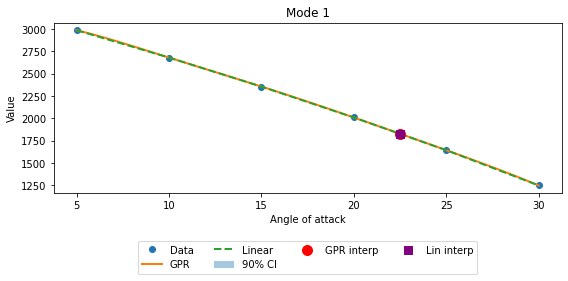}
        \caption{}
        \label{fig:Aoa_0}
    \end{subfigure}
    \hfill
    \begin{subfigure}[b]{0.48\textwidth}
        \centering
        \includegraphics[width=\textwidth]{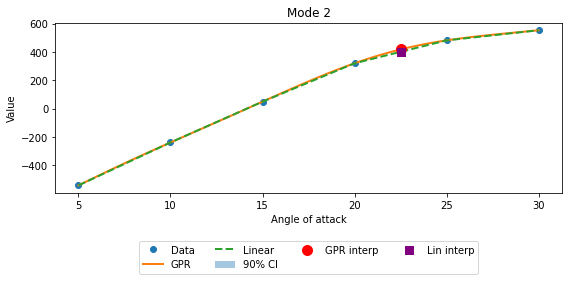}
        \caption{}
        \label{fig:Aoa_1}
    \end{subfigure}

    \vskip\baselineskip

    \begin{subfigure}[b]{0.48\textwidth}
        \centering
        \includegraphics[width=\textwidth]{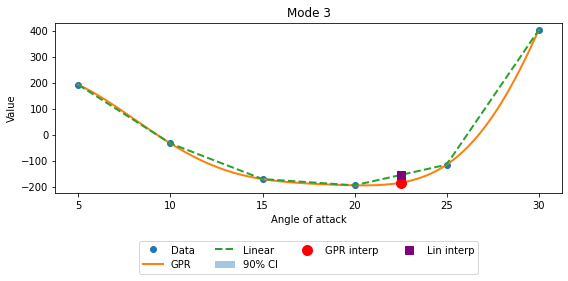}
        \caption{}
        \label{fig:AoA_2}
    \end{subfigure}
    \hfill
    \begin{subfigure}[b]{0.48\textwidth}
        \centering
        \includegraphics[width=\textwidth]{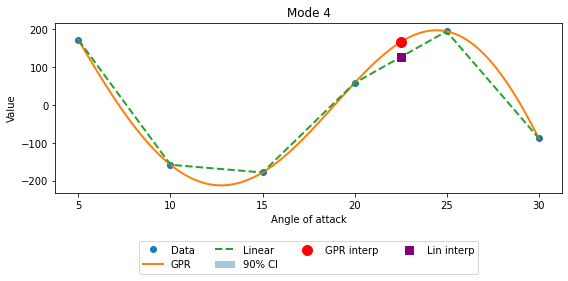}
        \caption{}
        \label{fig:AoA_3}
    \end{subfigure}

    \caption{Same as Fig. \ref{fig:Re_gpr} but for the AoA mode matrix.}
    \label{fig:Aoa_gpr}
\end{figure}

Figure~\ref{fig:Aoa_gpr} compares the results obtained by applying GPR and piecewise linear interpolation along each of the retained SVD modes associated with the AoA dimension. It can be observed that in the first SVD mode (Fig.~\ref{fig:Aoa_0}), the data distribution shows a linear trend with a negative slope, where both the regression and interpolation methods produce similar results. Similar to the first mode, the second mode shows initially a linear trend with a positive slope, where GPR and piecewise linear interpolation provide similar results up to the singular value corresponding to \(\text{AoA} = 20^{\circ}\), beyond which GPR smoothly captures the trend of the data. For the third and fourth modes, displayed in Fig.~\ref{fig:AoA_2} and Fig.~\ref{fig:AoA_3}, respectively, it can be observed that GPR addresses the non-linear behaviour of the data, providing a smooth fit.

\subsection{HOSVD analysis for the X and Y spatial dimensions}

\begin{figure}
    \centering
    \begin{subfigure}[b]{0.48\textwidth}
        \centering
        \includegraphics[width=\textwidth]{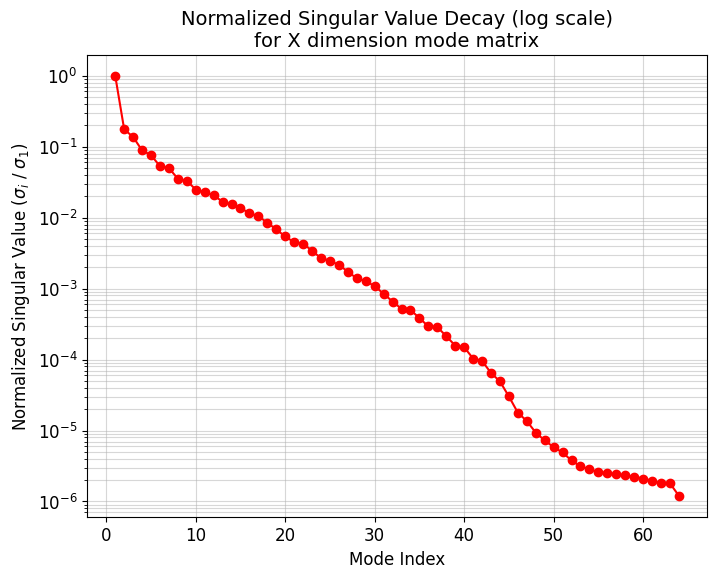}
        \caption{Singular value decay for the X dimension.}
        \label{fig:X_sv}
    \end{subfigure}
    \hfill
    \begin{subfigure}[b]{0.48\textwidth}
        \centering
        \includegraphics[width=\textwidth]{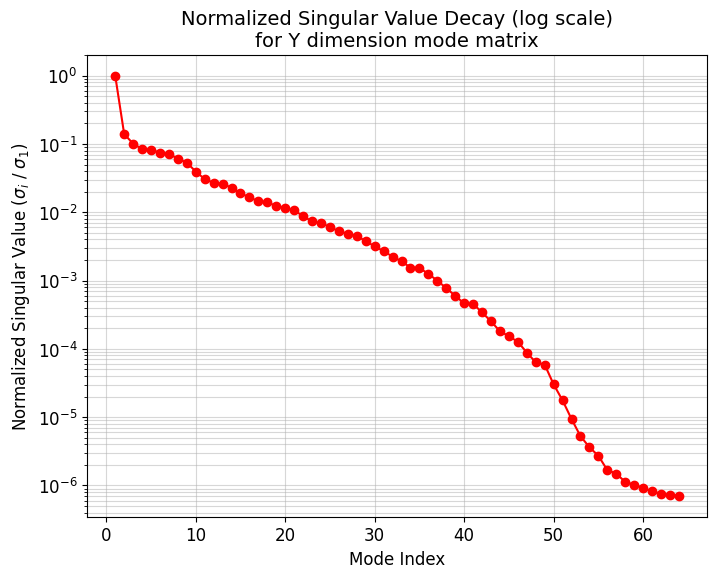}
        \caption{Singular value decay for the Y dimension.}
        \label{fig:Y_sv}
    \end{subfigure}

    \vspace{0.5em}
    \begin{subfigure}[b]{0.48\textwidth}
        \centering
        \includegraphics[width=\textwidth]{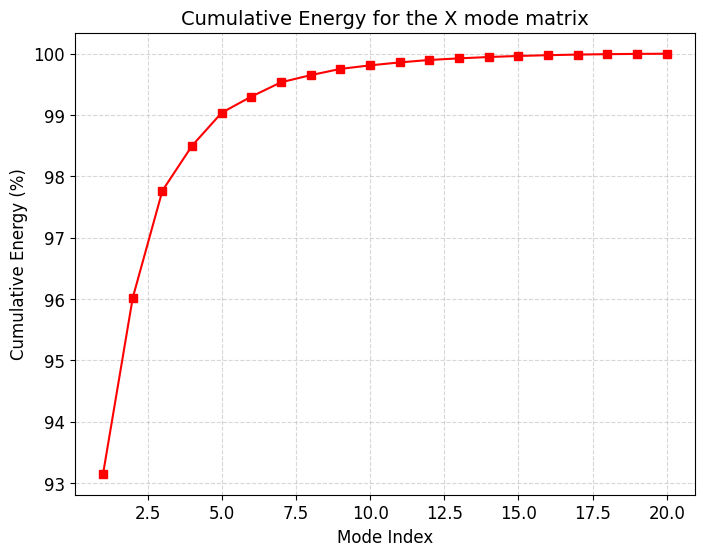}
        \caption{Cumulative energy (zoomed-in view) for the X dimension.}
        \label{fig:X_cum_zoom}
    \end{subfigure}
    \hfill
    \begin{subfigure}[b]{0.48\textwidth}
        \centering
        \includegraphics[width=\textwidth]{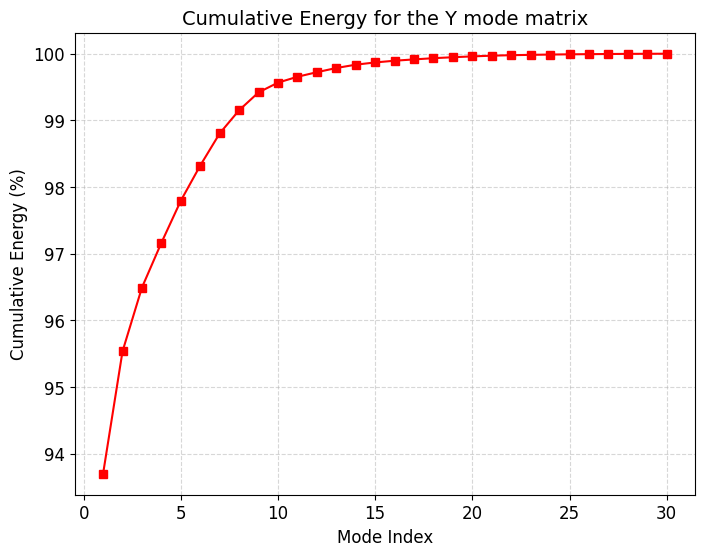}
        \caption{Cumulative energy (zoomed-in view) for the Y dimension.}
        \label{fig:Y_cum_zoom}
    \end{subfigure}

    \caption{Singular value decomposition (SVD) results for both X and Y dimensions. 
    (a,b) show the singular value decay, while (c,d) present the cumulative energy distribution (zoomed-in). 
    These plots illustrate how the dominant SVD modes capture the majority of the system’s variance in both spatial directions.}
    \label{fig:XY_SVD_combined}
\end{figure}

The singular value decay along the X and Y spatial dimensions, denoted as \(\sigma^{(4)}\) and \(\sigma^{(5)}\) respectively, is shown in Fig.~\ref{fig:Y_sv}. It is observed that, apart from the drop between the first and second singular values, there is no significant decay in magnitude thereafter. To determine an appropriate truncation threshold for dimensionality reduction, the cumulative energy distribution has also been plotted (Fig.~\ref{fig:Y_cum_zoom}). Figure~\ref{fig:X_cum_zoom} indicates that approximately 99.9\% of the total energy in the X dimension is captured by the first 11 SVD modes, while 15 modes are required along the Y dimension to retain 99.9\% of the energy (Fig.~\ref{fig:Y_cum_zoom}).

\subsection{HOSVD analysis and RNN forecasting for the temporal dimension}

\begin{figure}
    \centering
    \begin{subfigure}[b]{0.48\textwidth}
        \centering
        \includegraphics[width=\textwidth]{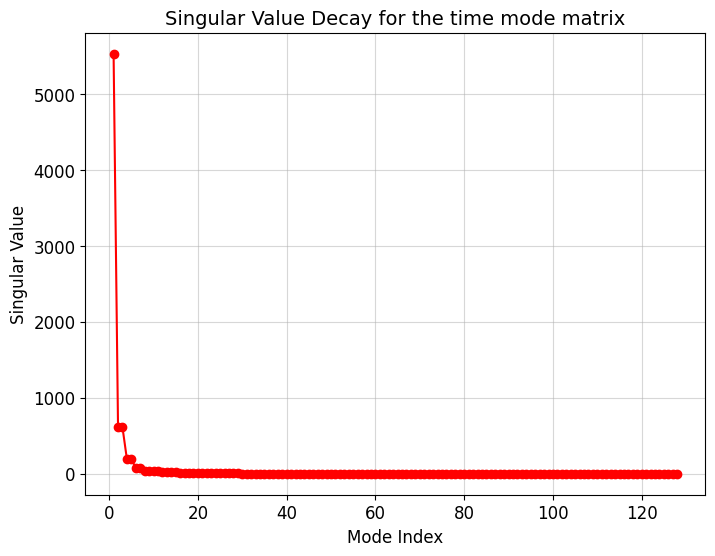}
        \caption{}
        \label{fig:T_sv}
    \end{subfigure}
    \hfill
    \begin{subfigure}[b]{0.48\textwidth}
        \centering
        \includegraphics[width=\textwidth]{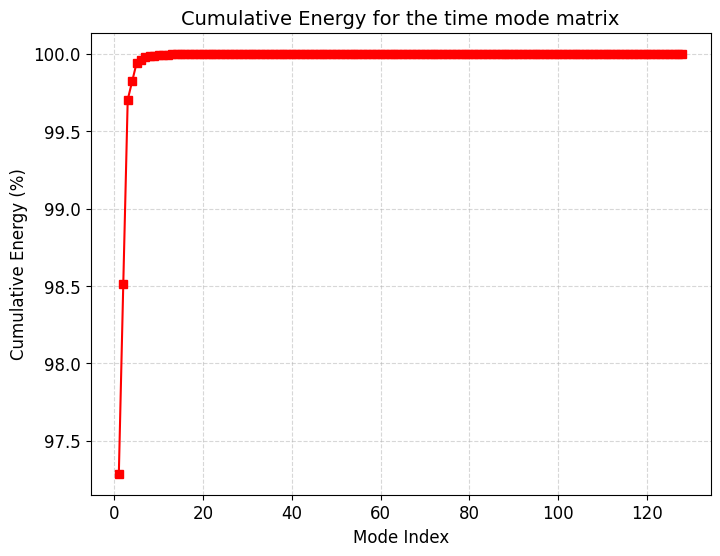}
        \caption{}
        \label{fig:T_cum}
    \end{subfigure}
    \caption{Same as Fig. \ref{fig:Re_sv}, but for the temporal SVD mode matrix.}
    \label{fig:T_sv1}
\end{figure}

Figure \ref{fig:T_sv1} displays the singular value decay and cumulative energy plots along the temporal dimension. Figure \ref{fig:T_sv} exhibits a significant decrease in the magnitudes of the first two SVD modes, followed by a more gradual decrease in subsequent modes. From the third to the fourth SVD mode, the decay becomes even less pronounced, and beyond this point, the singular values remain nearly constant, forming an almost horizontal line indicative of negligible decay. The cumulative energy plot displayed in Fig. \ref{fig:T_cum} shows that almost $99.9\%$ of the total energy is contained in the first 7 SVD modes. Thus, a highly accurate approximation of the full-order model can be achieved retaining 7 SVD modes.

\begin{figure}
\centering
\includegraphics[width=0.45\linewidth]{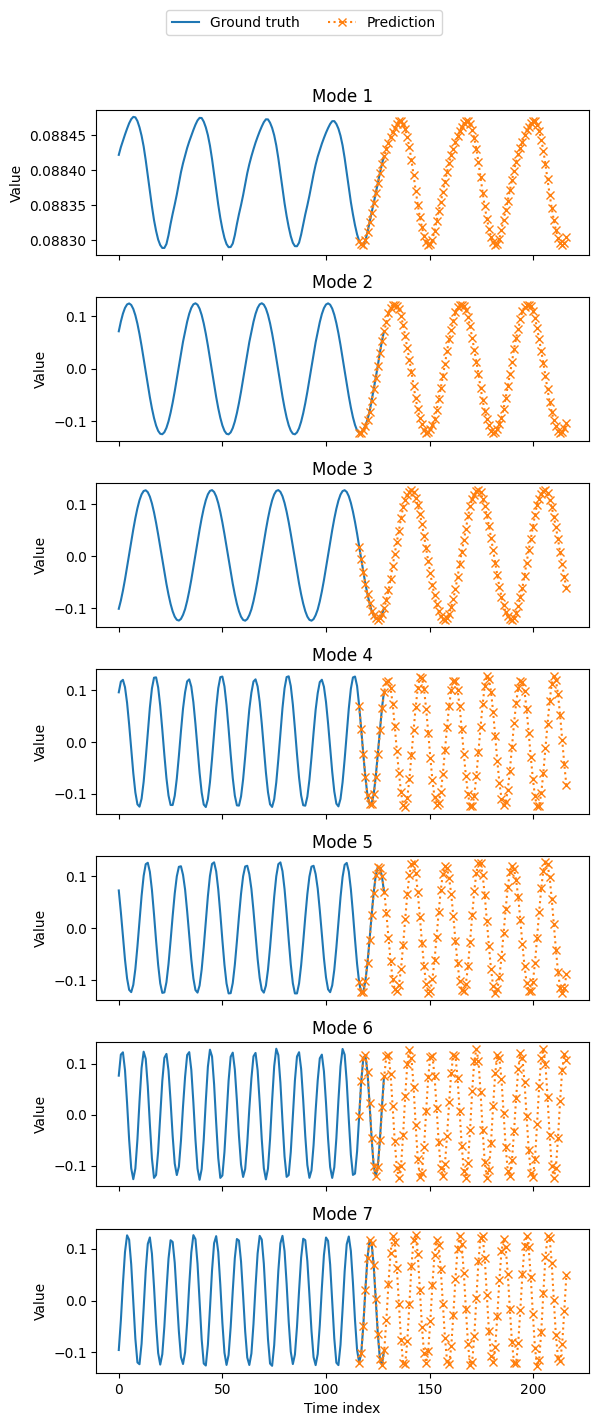}
\caption{ Forecast of the LSTM-based RNN on each of the 7 modes retained along the temporal dimension.} \label{prediction_modes}
\end{figure}

Figure~\ref{prediction_modes} presents the predictions obtained using the RNN for each of the retained SVD modes along the temporal mode matrix $\mathbf{T}$. The first three modes exhibit low-frequency oscillations, indicating the presence of dominant flow features. From the fourth mode onward, the frequency of oscillations increases, and slight variations in amplitude become more noticeable, reflecting more complex and higher-frequency dynamics. The region where the prediction overlaps with the ground truth corresponds to the test set. In this region, the predicted curves show a strong agreement with the ground truth, suggesting that the model generalizes well across the temporal domain. The smooth transition from ground truth to predicted data confirms that the RNN has effectively learned the underlying temporal dynamics of the system. Overall, the predicted time series matches the trends and frequency content of each mode, indicating that the neural network successfully captures the temporal behaviour embedded in the low-dimensional representation of the system.

\subsection{Filling missing data from incomplete databases}

\begin{figure}
    \centering
    \begin{subfigure}[b]{0.35\textwidth}
        \centering
        \includegraphics[width=\textwidth]{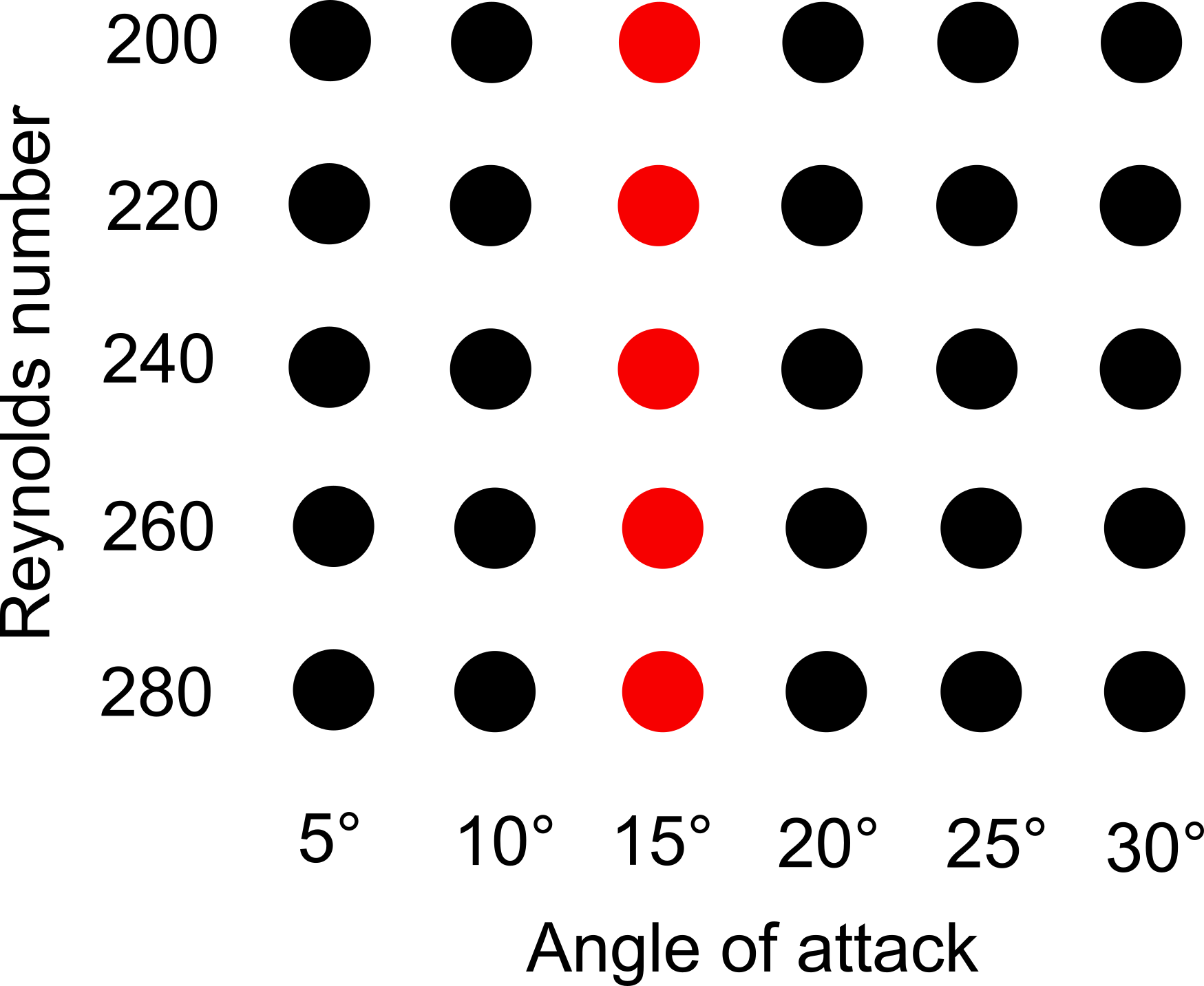}
        \caption{}
        \label{fig:F2_15}
    \end{subfigure}
    \hfill
    \begin{subfigure}[b]{0.35\textwidth}
        \centering
        \includegraphics[width=\textwidth]{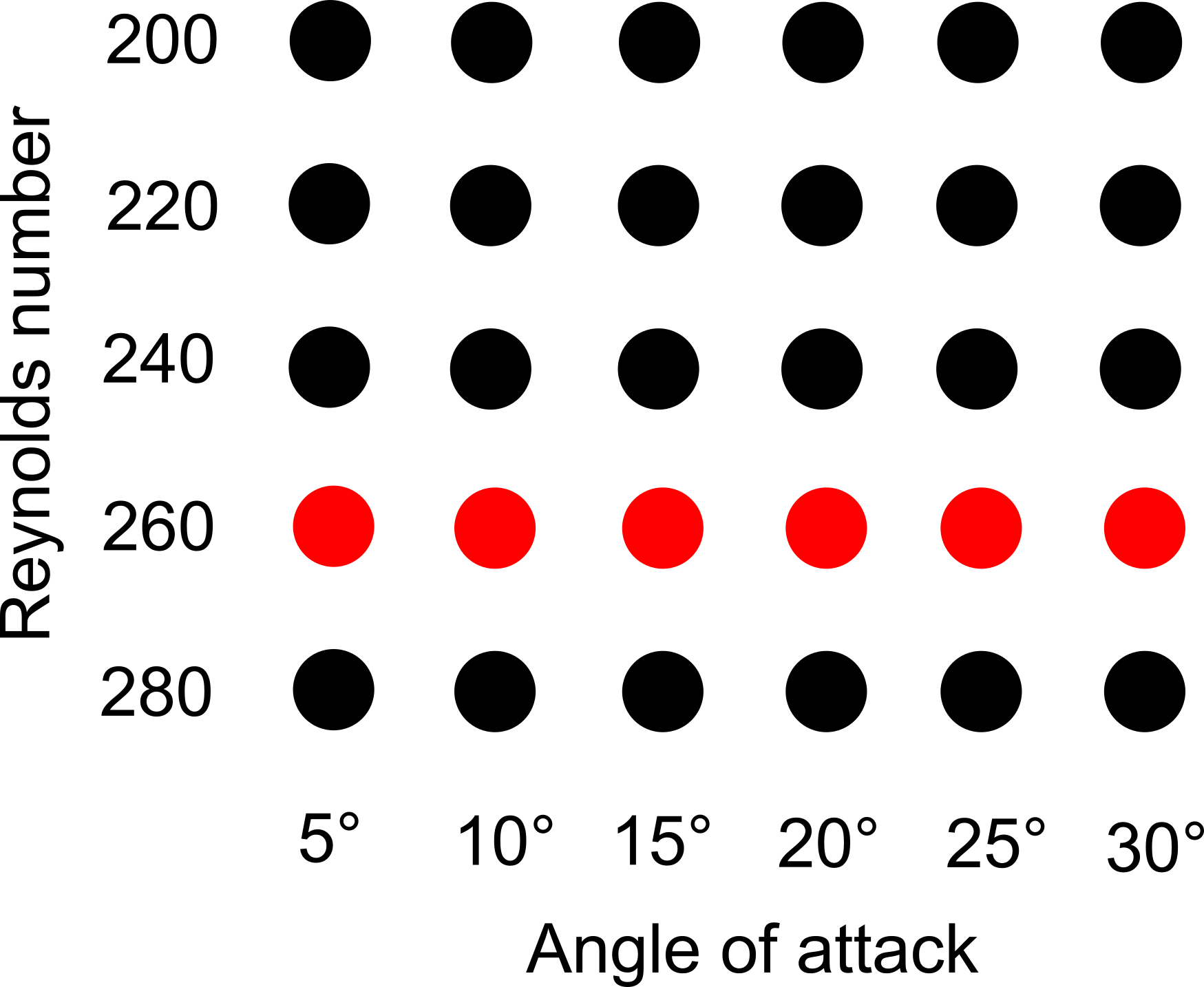}
        \caption{}
        \label{fig:F1_240}
    \end{subfigure}
    \caption{ Design space (in black) and interpolated data (in red) for the filling missing data from incomplete databases scenario. (a) Generation of a new database for AoA = $15^\circ$ and all the Re numbers (FAoA task), (a) Generation of a new database for Re = $260$ and all the AoA (FRe task). }
    \label{fig:DSF}
\end{figure}

This section presents the results obtained using the proposed MoTIF to fill in missing flow data from incomplete databases~(FRe, FAoA tasks on Tab. \ref{tab: testcases}), where one or more flow conditions are missing from the original database. Figure \ref{fig:F1_240} summarizes the two different cases considered, where the goal is to obtain an evenly sampled database in terms of Re and AoA: filling in a missing AoA for all Re numbers (Fig. \ref{fig:F2_15}), and filling in a missing Re number for all AoA values. Table \ref{tab:rrmse} collects the RRMSE for streamwise and normal velocity components for the different test cases addressed using MoTIF.

\begin{figure}
\centering
\includegraphics[width=1\linewidth]{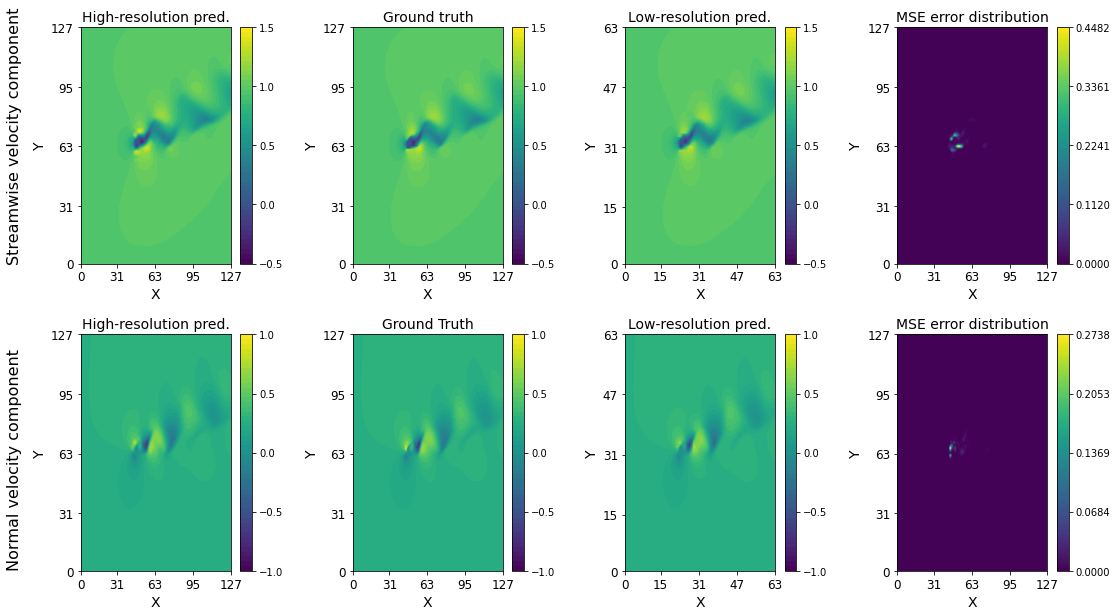}
\caption{ FAoA task; High-resolution prediction results obtained using the proposed MoTIF, with Re = 240 and AoA = $15^{\circ}$. From left to right and top to bottom: high-resolution prediction, ground truth data, low-resolution prediction and MSE of the streamwise and normal velocity component, for the highest MSE snapshot of each velocity component.} \label{F2_mse}
\end{figure}

Figure~\ref{F2_mse} presents a comparison of the predictions obtained using the proposed MoTIF model to reconstruct missing data in the database for the FAoA case (AoA = $15^\circ$). The model predicts the flow fields across all Re in the database for this missing AoA. Although the interpolation is performed along the AoA dimension with one value missing, the model accurately captures the flow behaviour. The streamwise and normal velocity components are well predicted in both the high-resolution case (with an upscaling factor of 2) and the low-resolution one, closely matching the ground truth data. The MSE distribution for both velocity components shows the highest values near the square cylinder, which is mainly due to the dimensionality reduction. However, in the wake region and the rest of the domain, the MSE values are very low, indicating good predictive performance overall.

\begin{figure}
\centering
\includegraphics[width=1\linewidth]{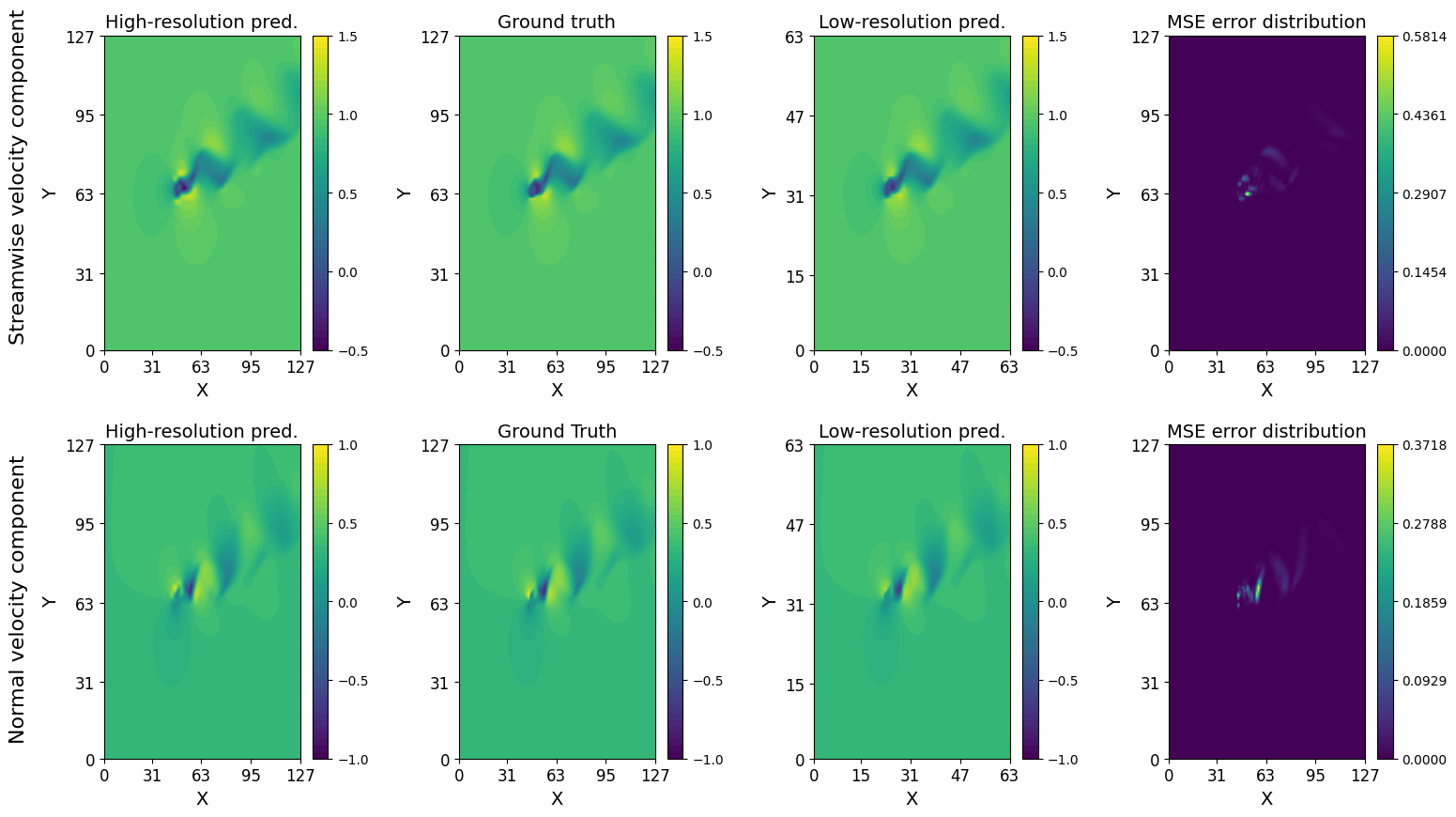}
\caption{FRe task; Same as \ref{F2_mse} but with Re = 260 and AoA = $20^\circ$.} \label{F1_mse}
\end{figure}

Figure~\ref{F1_mse} shows the streamwise and normal velocity contours predicted for the FRe case (Re= 260). The model predicts the flow fields across all AoA in the database for this missing Re. Similar to Fig.~\ref{F2_mse}, the interpolation is performed along an incomplete set of Reynolds numbers and demonstrates high accuracy in predicting the flow field. For both velocity components, the MoTIF model successfully reconstructs and upscales the flow structures both upstream and downstream of the square cylinder, capturing the flow behaviour under these specific conditions. The MSE distribution for the streamwise velocity shows the highest values near the top and bottom of the bluff body. In contrast, for the normal velocity component, the maximum MSE values are located in the wake region behind the cylinder.

\begin{figure}
\centering
\includegraphics[width=1\linewidth]{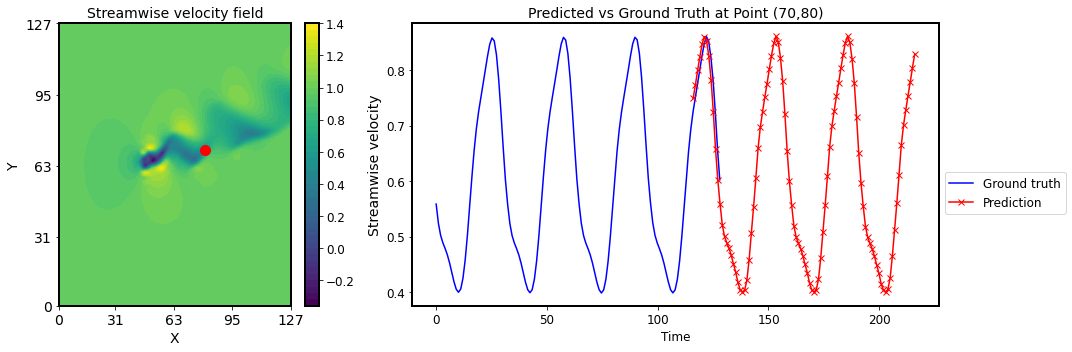}
\caption{FAoA task (Re = 240 and AoA = $15^\circ$); Temporal evolution of the streamwise velocity component in a point of interest of the predicted database.} \label{F2_time}
\end{figure}

\begin{figure}
\centering
\includegraphics[width=1\linewidth]{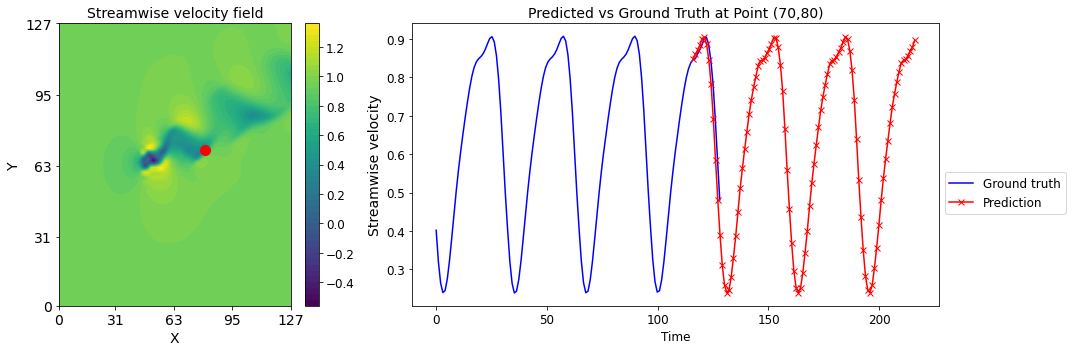}
\caption{FRe task; Same as \ref{F2_time} but with Re = 260 and AoA  = $20^\circ$.} \label{F1_time}
\end{figure}

The evolution of the streamwise velocity at a point of interest $(x = 70, y = 80)$, located in the wake region behind the bluff body, is shown in Fig.~\ref{F2_time} for the FRe case and Fig.~\ref{F1_time} for the FAoA case. The predicted and reference values closely match over time steps 106 to 128, which correspond to the test set excluded during the training of the LSTM-based neural network. This overlap indicates that the network was properly trained using the 7 retained SVD modes from the temporal mode matrix $\mathbf{T}$ and is capable of accurately predicting future flow snapshots. Beyond the test interval, the predicted time series continues to follow the ground truth data closely, confirming that the model effectively captures the temporal dynamics of the flow at this location.

\begin{figure}
    \centering
    \begin{subfigure}[b]{0.48\textwidth}
        \centering
        \includegraphics[width=\textwidth]{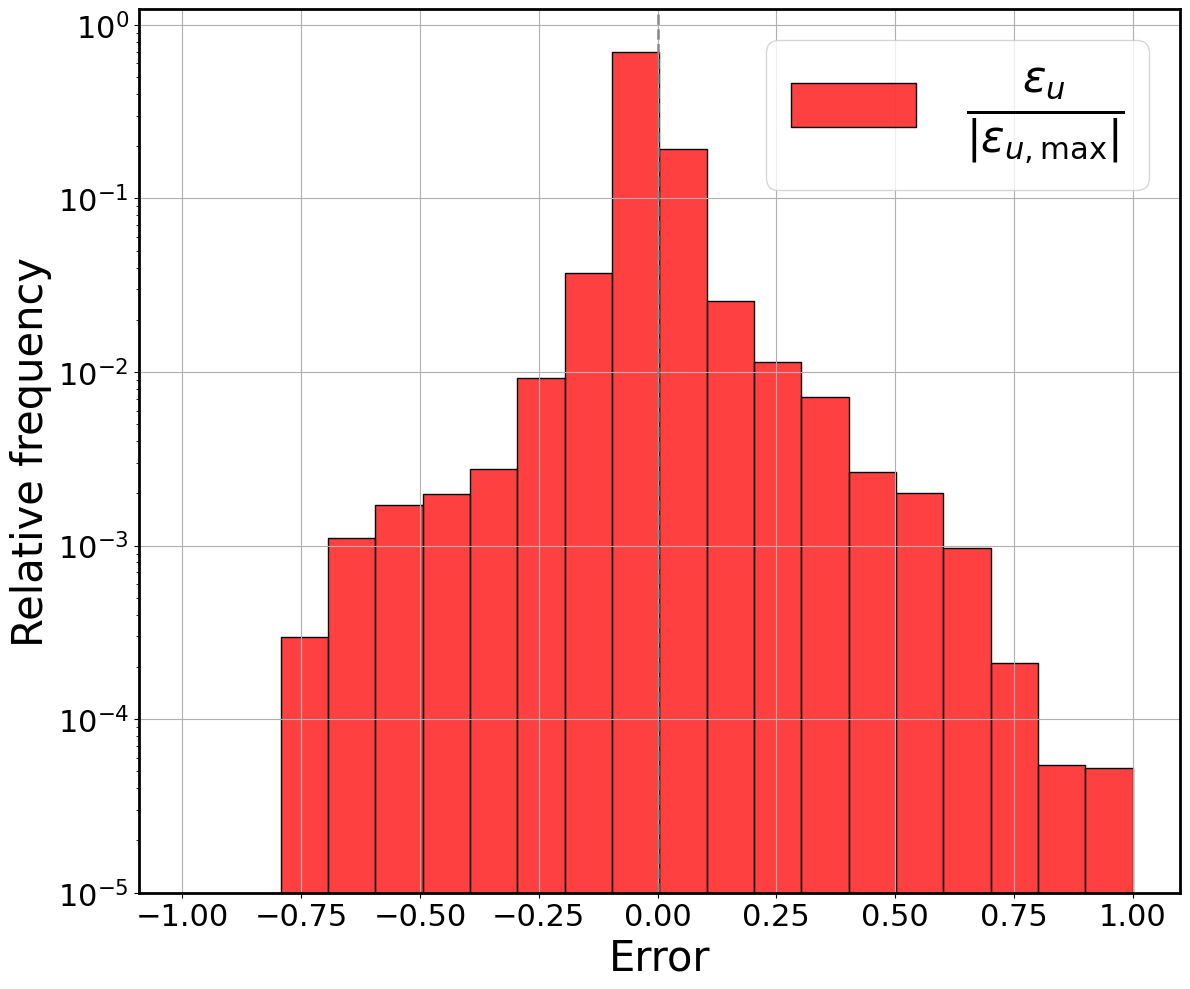}
        \caption{}
        \label{fig:F2_st}
    \end{subfigure}
    \hfill
    \begin{subfigure}[b]{0.48\textwidth}
        \centering
        \includegraphics[width=\textwidth]{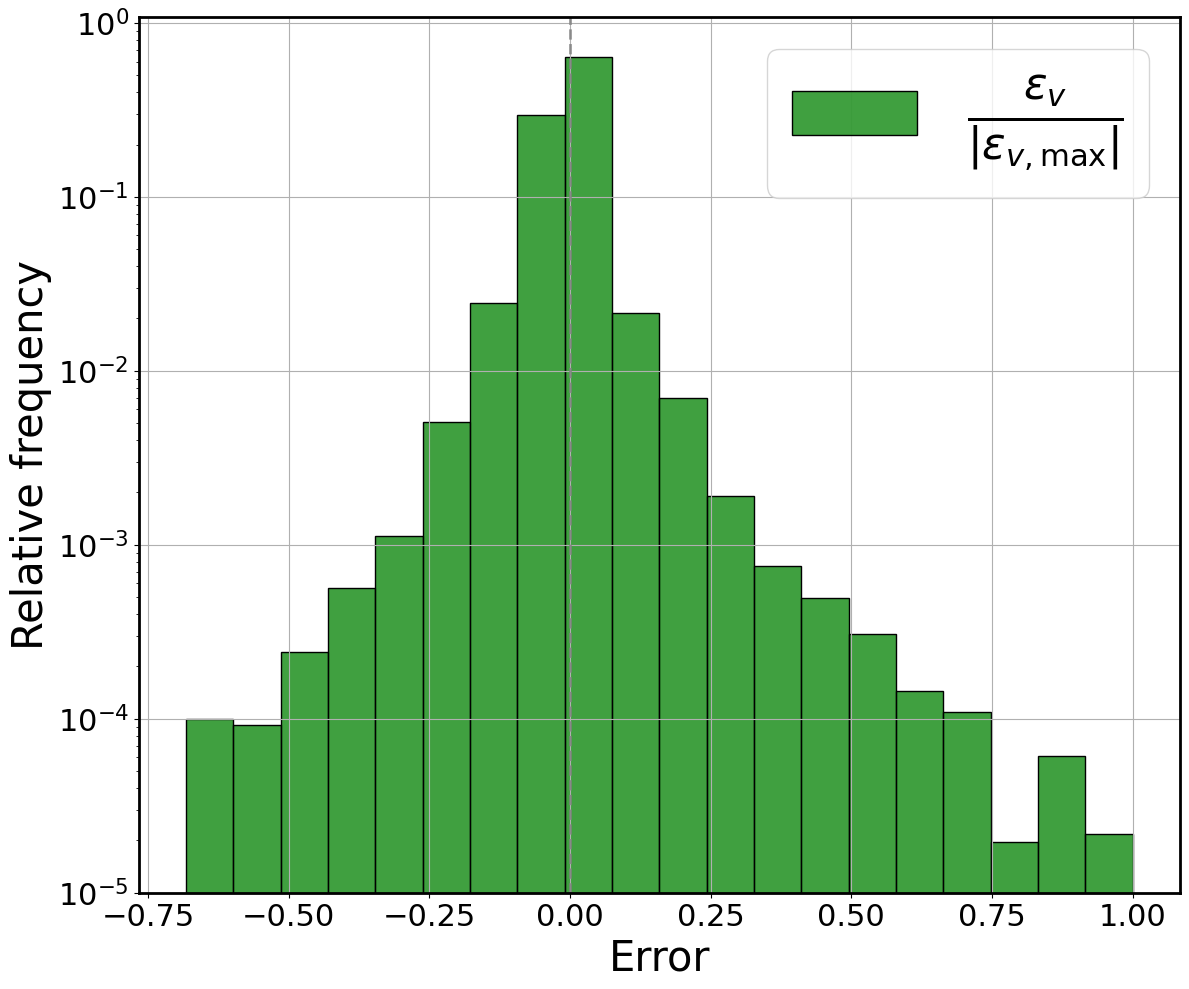}
        \caption{}
        \label{fig:F2_norm}
    \end{subfigure}
    \caption{Fill inconsistent databases scenario, FRe test case (Re = 240, AoA = $15^{\circ}$): normalized error relative frequency histograms. (a) streamwise velocity, (b)normal velocity. The histogram has been built with 21 bins.}
    \label{fig:F2_relfreq}
\end{figure}

Figure~\ref{fig:F2_relfreq} shows the normalized error distribution plotted as relative frequency histograms for the streamwise and normal velocity components in the FRe test case. Figure~\ref{fig:F2_st} indicates that almost $90\%$ of the error values for the streamwise component are clustered near zero and a slight bias toward negative values. The normalized error histogram for the normal velocity component (Fig.~\ref{fig:F2_st}) shows that over $90\%$ of the values centred around zero within an error range of $\pm 0.1$, showing no visible bias toward positive or negative values. This distribution aligns well with the MSE distribution presented in Fig.~\ref{F2_mse} and the low RRMSE values detailed in Tab.~\ref{tab:rrmse}.

\begin{figure}
    \centering
    \begin{subfigure}[b]{0.48\textwidth}
        \centering
        \includegraphics[width=\textwidth]{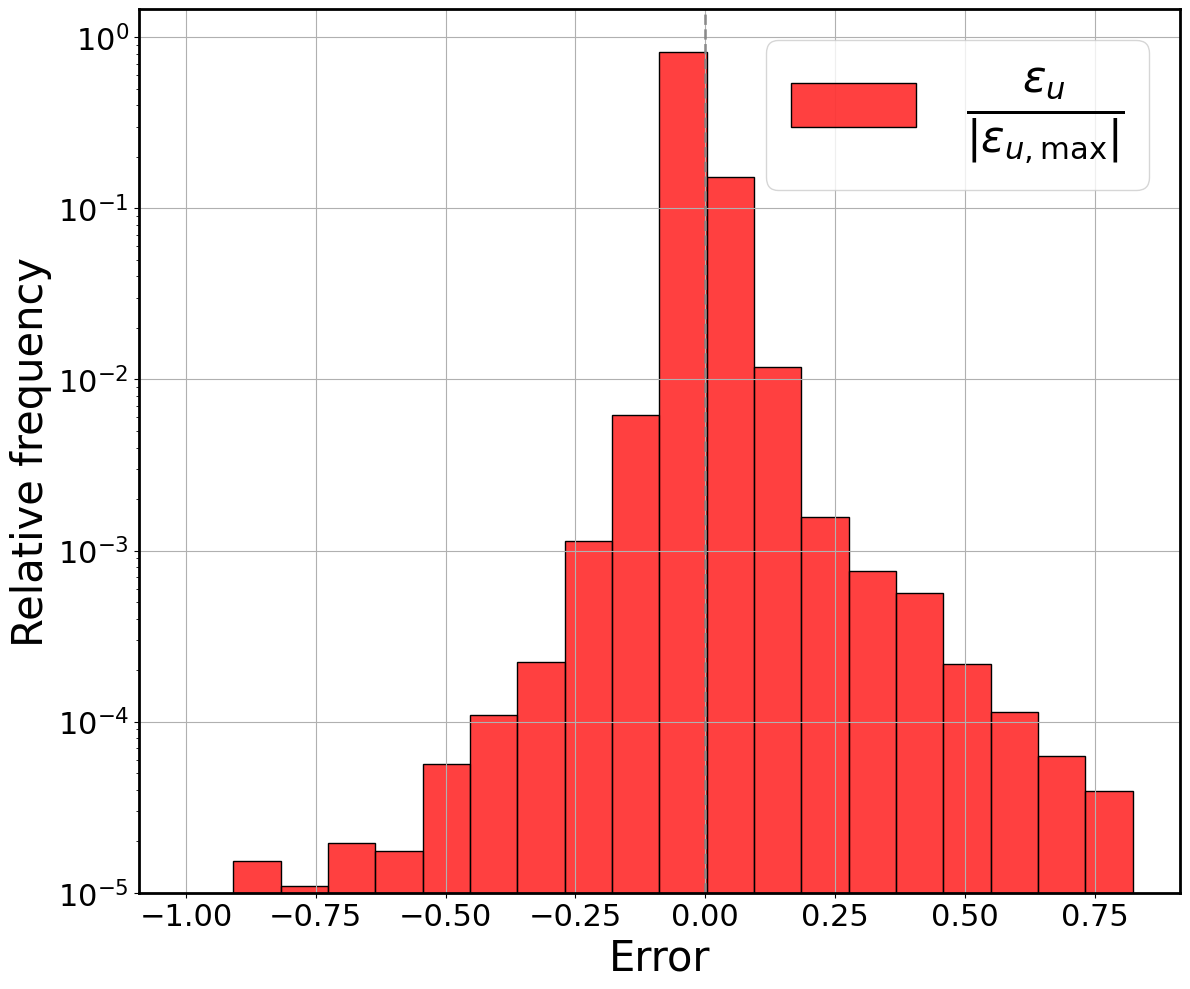}
        \caption{}
        \label{fig:F1_st}
    \end{subfigure}
    \hfill
    \begin{subfigure}[b]{0.48\textwidth}
        \centering
        \includegraphics[width=\textwidth]{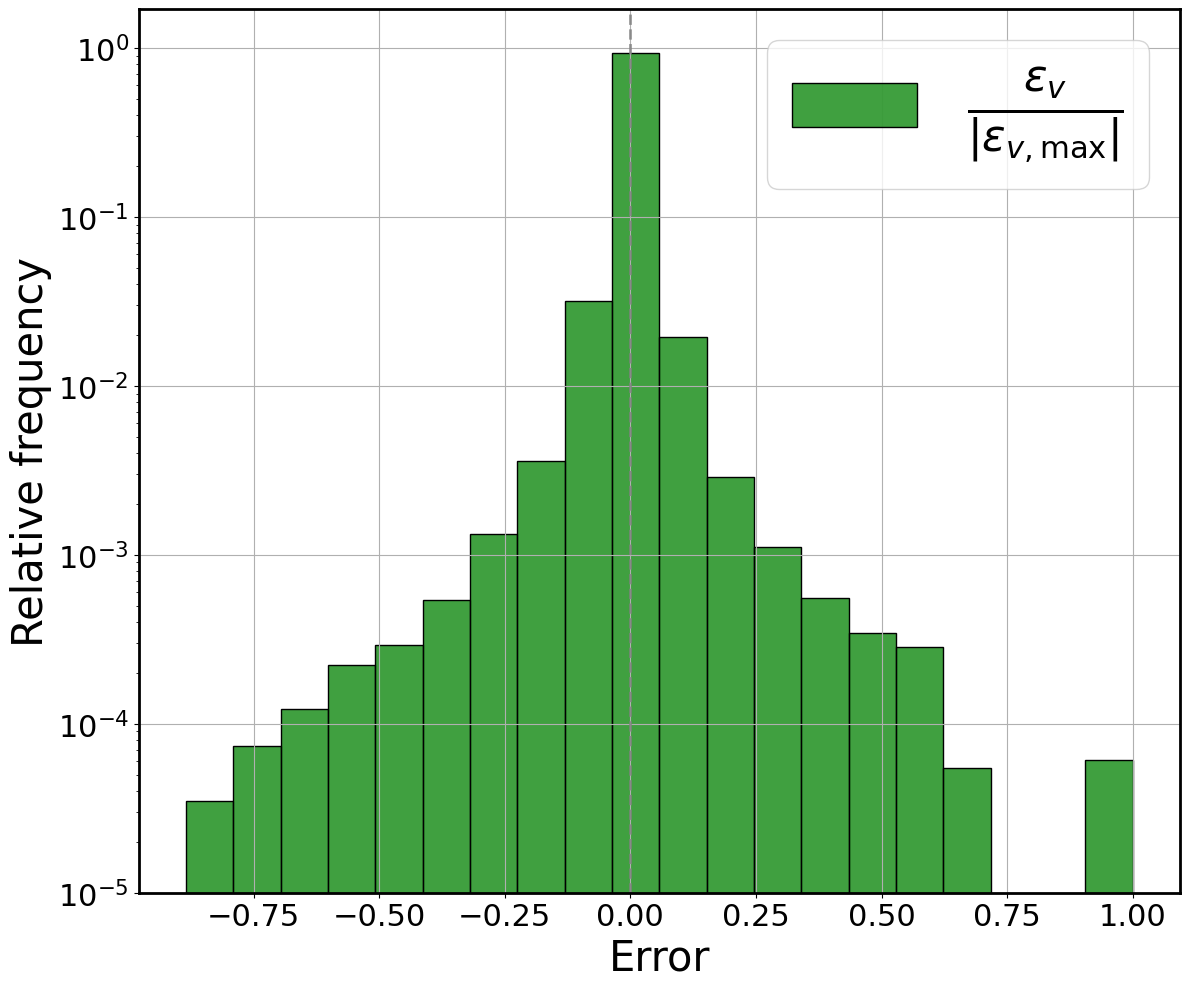}
        \caption{}
        \label{fig:F1_norm}
    \end{subfigure}
    \caption{Same as \ref{fig:F2_relfreq} but for the FRe case, with Re = 260 and AoA = $20^\circ$.}
    \label{fig:F1_relfreq}
\end{figure}

In the same manner as Fig.~\ref{fig:F2_relfreq}, the normalized error relative frequency for the streamwise and normal velocity components in the FAoA test case is shown in Fig.~\ref{fig:F1_relfreq}. Similar to the FRe case, the error distributions for both components show that over $90\%$ of the values clustered around zero. The distribution corresponding to the streamwise velocity component shows a slight bias toward negative values. These distributions are consistent with the low prediction RRMSE reported in Tab.~\ref{tab:rrmse} and the MSE distribution detailed above.

\subsection{Generation of databases for unseen flow conditions.}

\begin{figure}[H]
    \centering
    \begin{subfigure}[b]{0.35\textwidth}
        \centering
        \includegraphics[width=\textwidth]{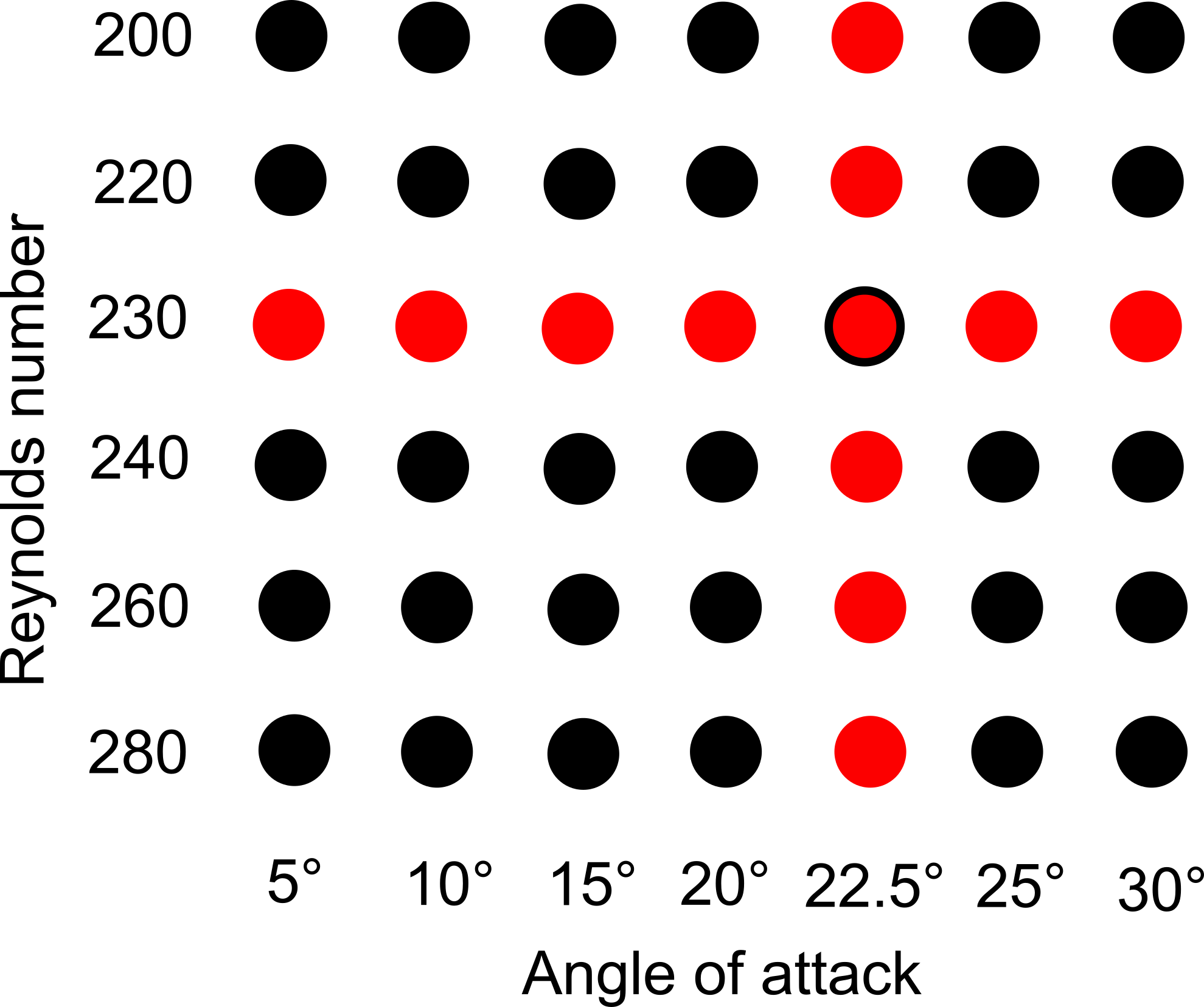}
        \caption{}
        \label{fig:N1}
    \end{subfigure}
    \hfill
    \begin{subfigure}[b]{0.35\textwidth}
        \centering
        \includegraphics[width=\textwidth]{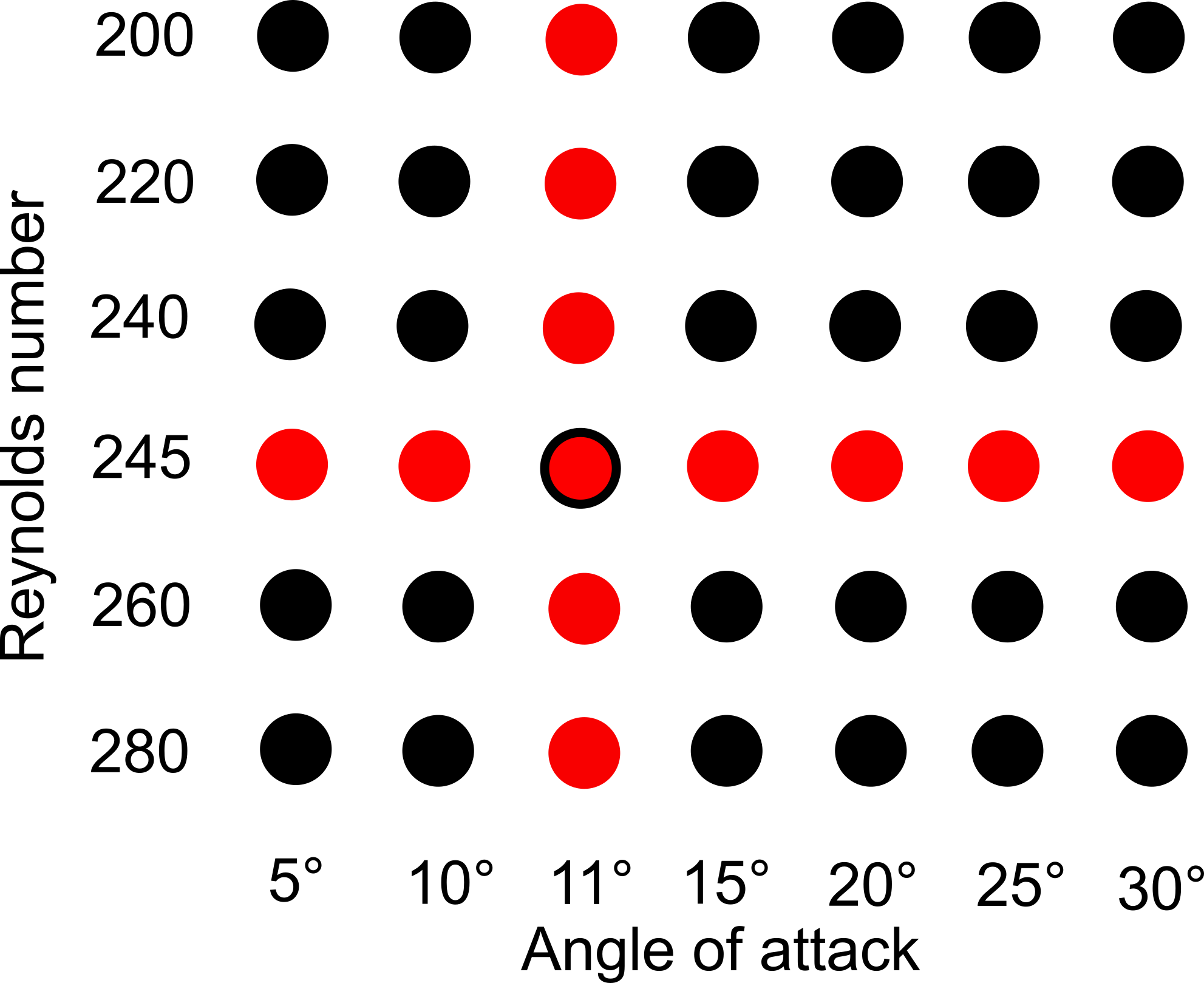}
        \caption{}
        \label{fig:N2}
    \end{subfigure}
    \caption{ Same as Fig. \ref{fig:DSF} but for the generation of new databases for unseen flow conditions scenario. The dots in red refer to the added unseen flow conditions, while the highlighted ones in black refer to the target flow condition.}
    \label{fig:DSN}
\end{figure}

This section presents the results obtained for generating data under unseen flow conditions using the proposed MoTIF. Two distinct cases were considered, with the objective of predicting flow conditions where neither the AoA nor the Reynolds number exist in the original database: generating a database for AoA = $22.5^\circ$ at Re = 230, referred to as N1 (Fig.~\ref{fig:N1}), and for AoA = $11^\circ$ at Re = 245, referred to as N2 (Fig.~\ref{fig:N2}). These test cases were strategically chosen to evaluate the capabilities of the proposed methodology: the N1 task lies between two equispaced entries in the original database, while the N2 task corresponds to a prediction near an existing database value.

\begin{figure}
\centering
\includegraphics[width=1\linewidth]{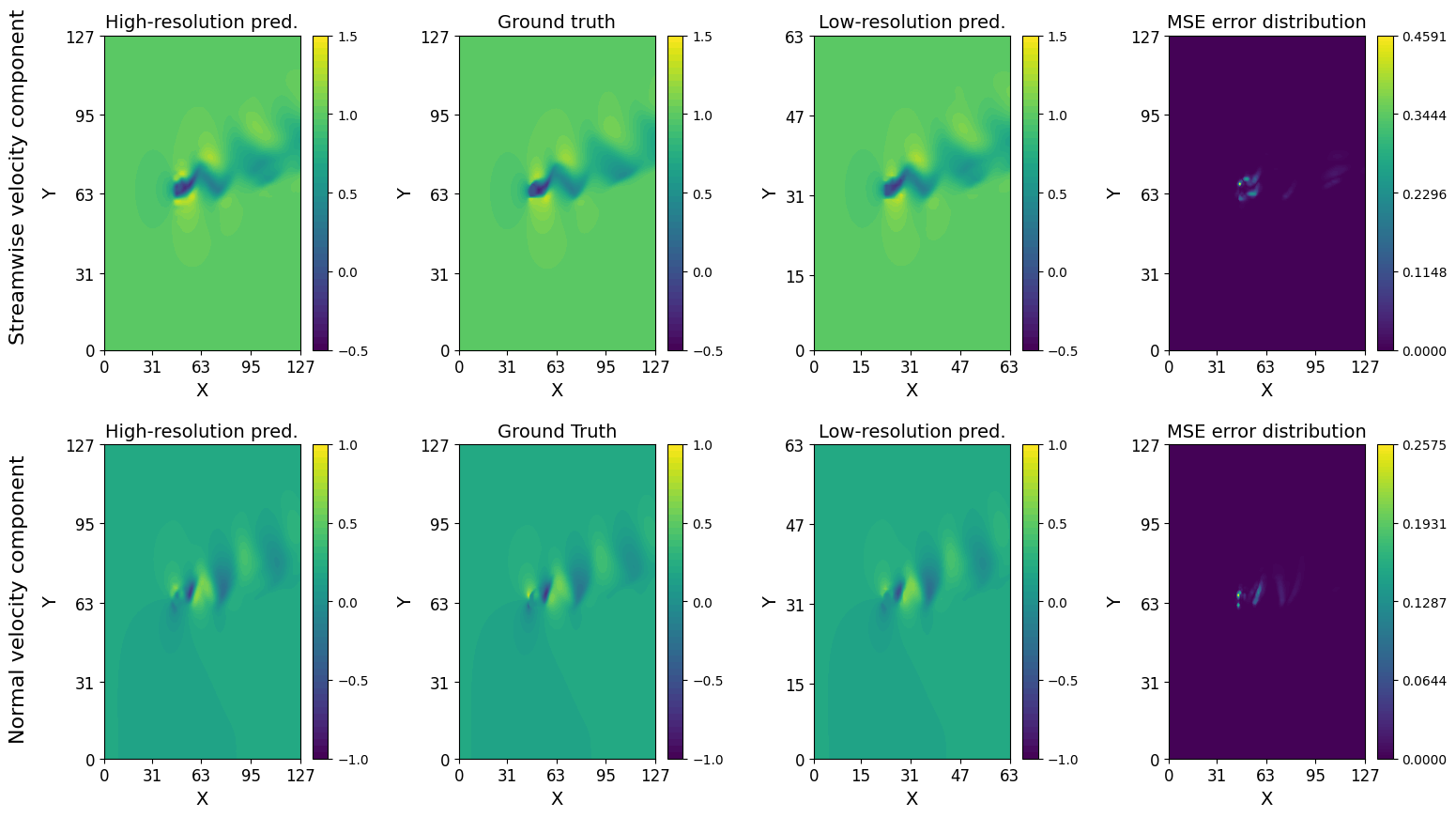}
\caption{N1 task; Same as \ref{F2_mse} but with Re = 230 and AoA = $22.5^\circ$.} \label{N1_mse}
\end{figure}

Figure~\ref{N1_mse} displays the predicted streamwise and normal velocity components for the N1 case. It is evident that, for both velocity components, the high-resolution predictions accurately capture the flow phenomena, closely matching the structures observed in the ground truth. Similar to the filled-in data for incomplete databases, the MSE distribution remains close to zero over most of the domain, with maximum errors occurring near the top and bottom of the bluff body for the streamwise velocity, and in a localized region downstream of the bluff body for the normal velocity component.
   
\begin{figure}
\centering
\includegraphics[width=1\linewidth]{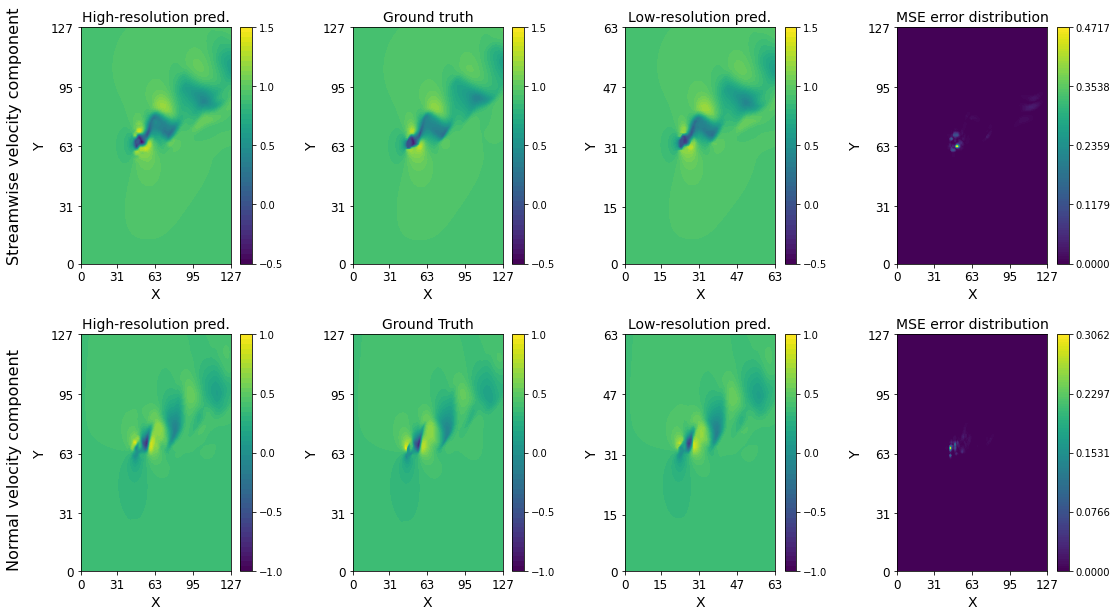}
\caption{N2 task; Same as \ref{F2_mse} but with Re = 245 and AoA = $11^\circ$.} \label{N2_mse}
\end{figure}

The predictions obtained for the N2 test case are shown in Fig.~\ref{N2_mse}. Similar to the N1 case, the proposed MoTIF accurately captures the fluid dynamics phenomena. The flow structures for both the streamwise and normal velocity components are precisely reconstructed in both the low- and high-resolution predictions, demonstrating the ability of GPR to spatially upscale with high fidelity to the original data. Regarding the MSE distribution, the maximum errors occur in the vicinity of the bluff body for both velocity components.

\begin{figure}
\centering
\includegraphics[width=1\linewidth]{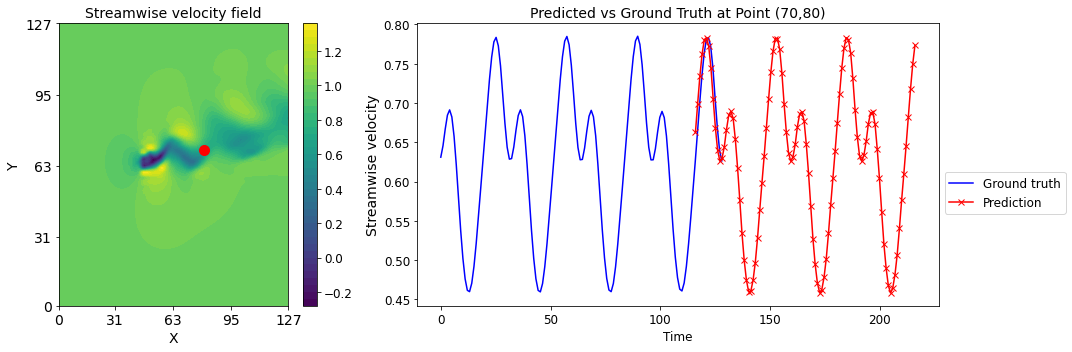}
\caption{N1 task; Same as \ref{F2_time} but with Re = 230 and AoA = $22.5^\circ$.} \label{N1_time}
\end{figure}

\begin{figure}
\centering
\includegraphics[width=1\linewidth]{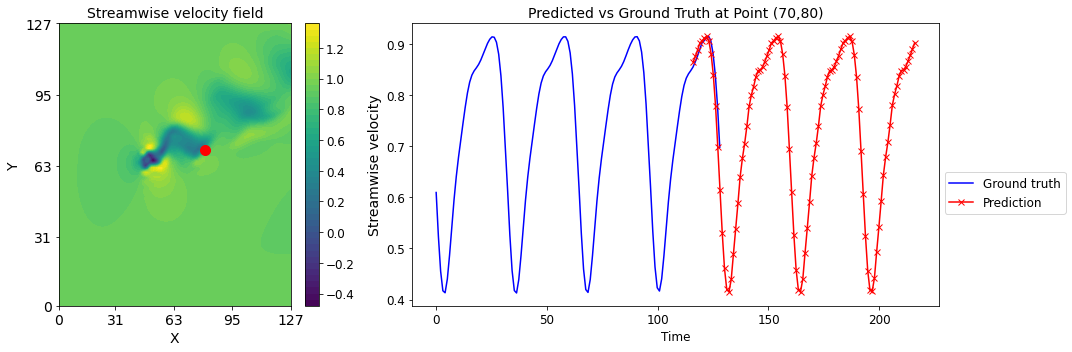}
\caption{N2 task; Same as \ref{F2_mse} but  with Re = 245 and AoA = $11^\circ$.} \label{N2_time}
\end{figure}

Figures~\ref{N1_time} and~\ref{N2_time} show the evolution of the streamwise velocity magnitude at a point of interest $(x = 70, y = 80)$, located within the wake behind the square cylinder. For both test cases, the original and predicted time series overlap closely over the snapshots corresponding to the test set (between snapshots 106 and 128), indicating that the LSTM-based neural network was adequately trained on the temporal coefficients associated with the retained temporal SVD modes.

\begin{figure}
    \centering
    \begin{subfigure}[b]{0.48\textwidth}
        \centering
        \includegraphics[width=\textwidth]{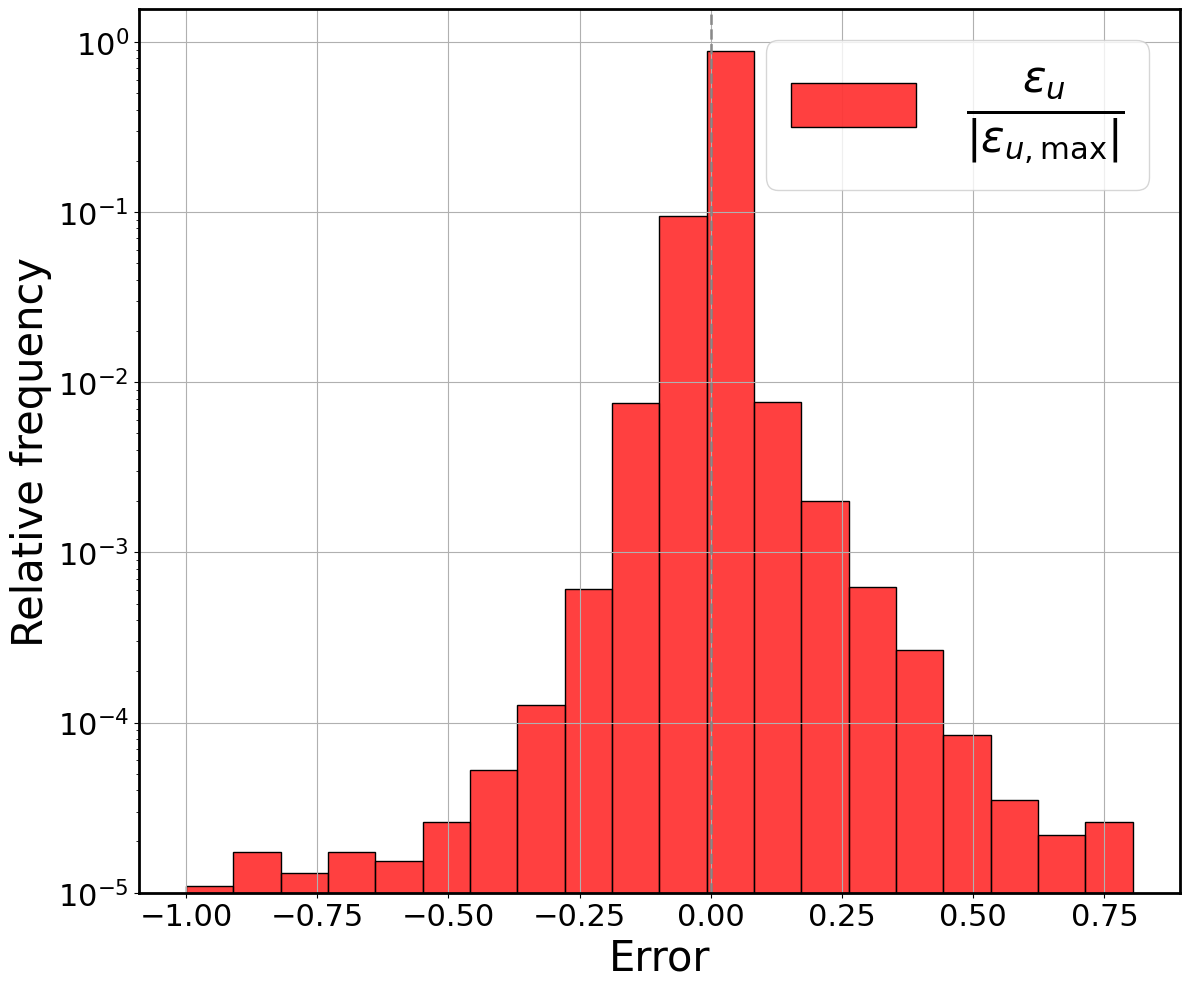}
        \caption{}
        \label{fig:N1_st}
    \end{subfigure}
    \hfill
    \begin{subfigure}[b]{0.48\textwidth}
        \centering
        \includegraphics[width=\textwidth]{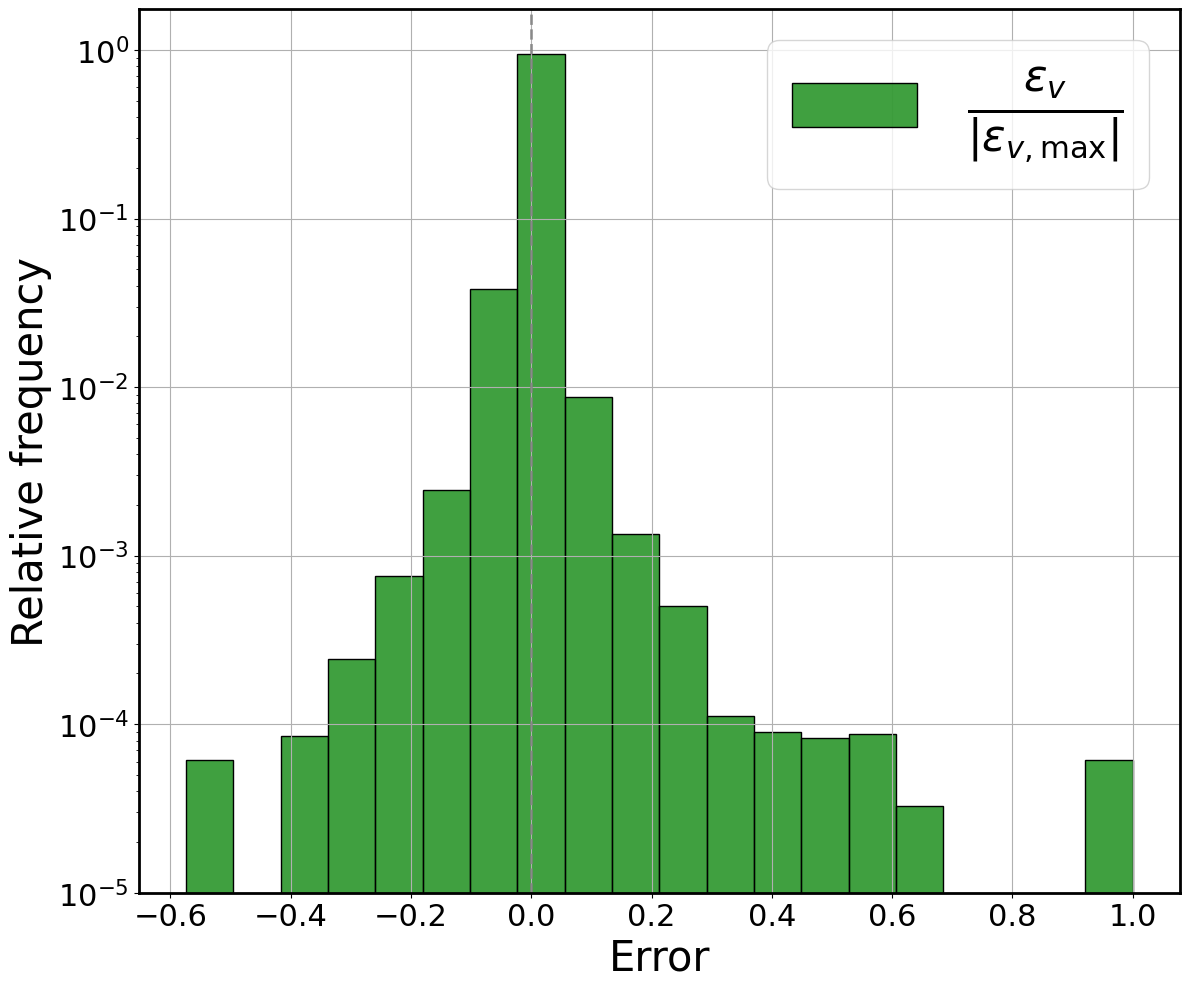}
        \caption{}
        \label{fig:N1_norm}
    \end{subfigure}
    \caption{N1 task; Same as \ref{fig:F2_relfreq} but with Re = 230 and AoA = $22.5^\circ$.}
    \label{fig:N1_relfreq}
\end{figure}

Figure~\ref{fig:N1_relfreq} presents the normalized error distribution for the N1 test case, shown as relative frequency histograms for the streamwise and normal velocity components. Figures~\ref{fig:N1_st} and~\ref{fig:N1_norm} indicate that over $90\%$ of the error values for both components are concentrated around zero, with a slight bias toward negative values. This behavior is consistent with the MSE distribution observed in Fig.~\ref{N1_mse} and the low RRMSE values reported in Tab.~\ref{tab:rrmse}.

\begin{figure}
    \centering
    \begin{subfigure}[b]{0.48\textwidth}
        \centering
        \includegraphics[width=\textwidth]{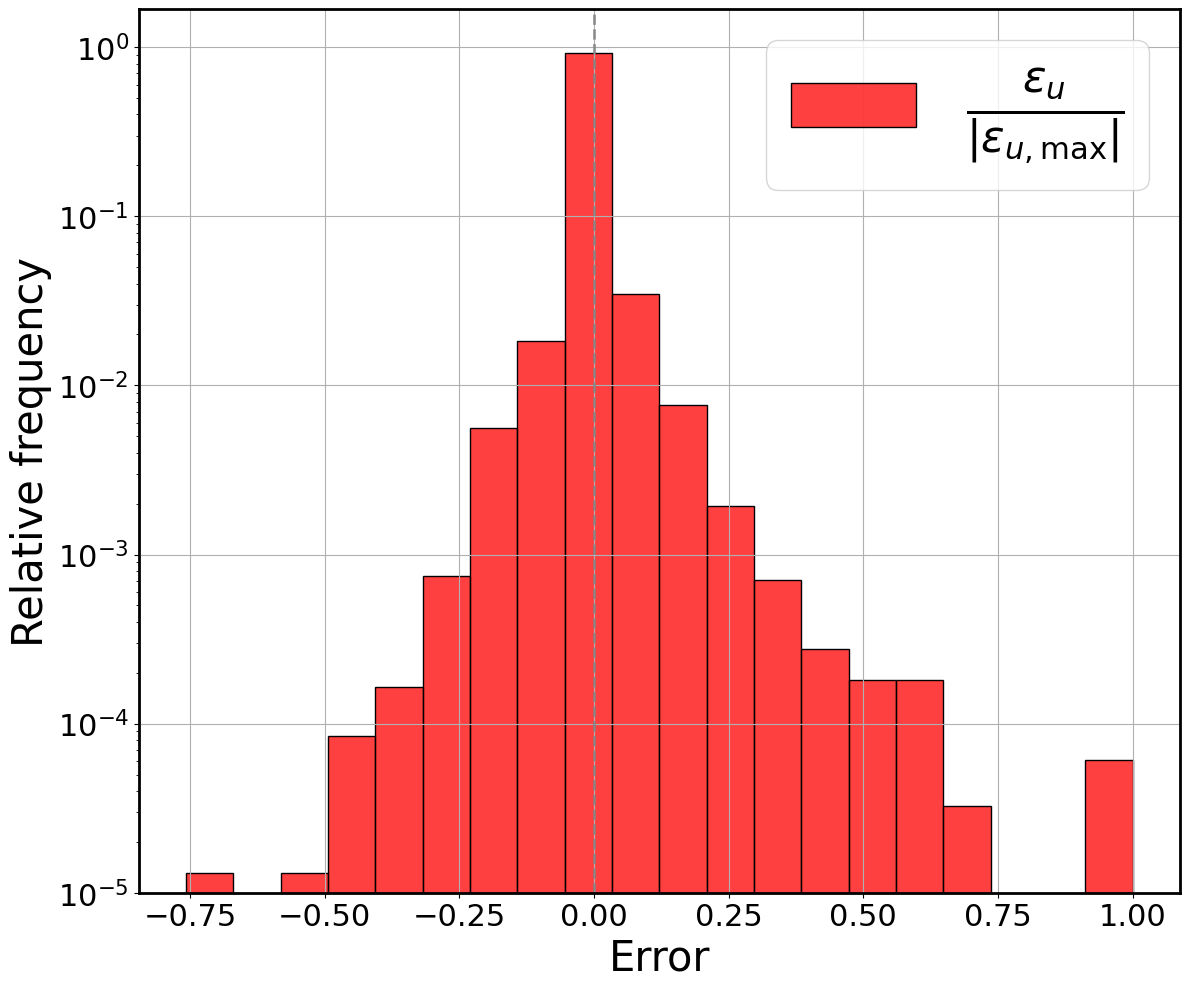}
        \caption{}
        \label{fig:N2_st}
    \end{subfigure}
    \hfill
    \begin{subfigure}[b]{0.48\textwidth}
        \centering
        \includegraphics[width=\textwidth]{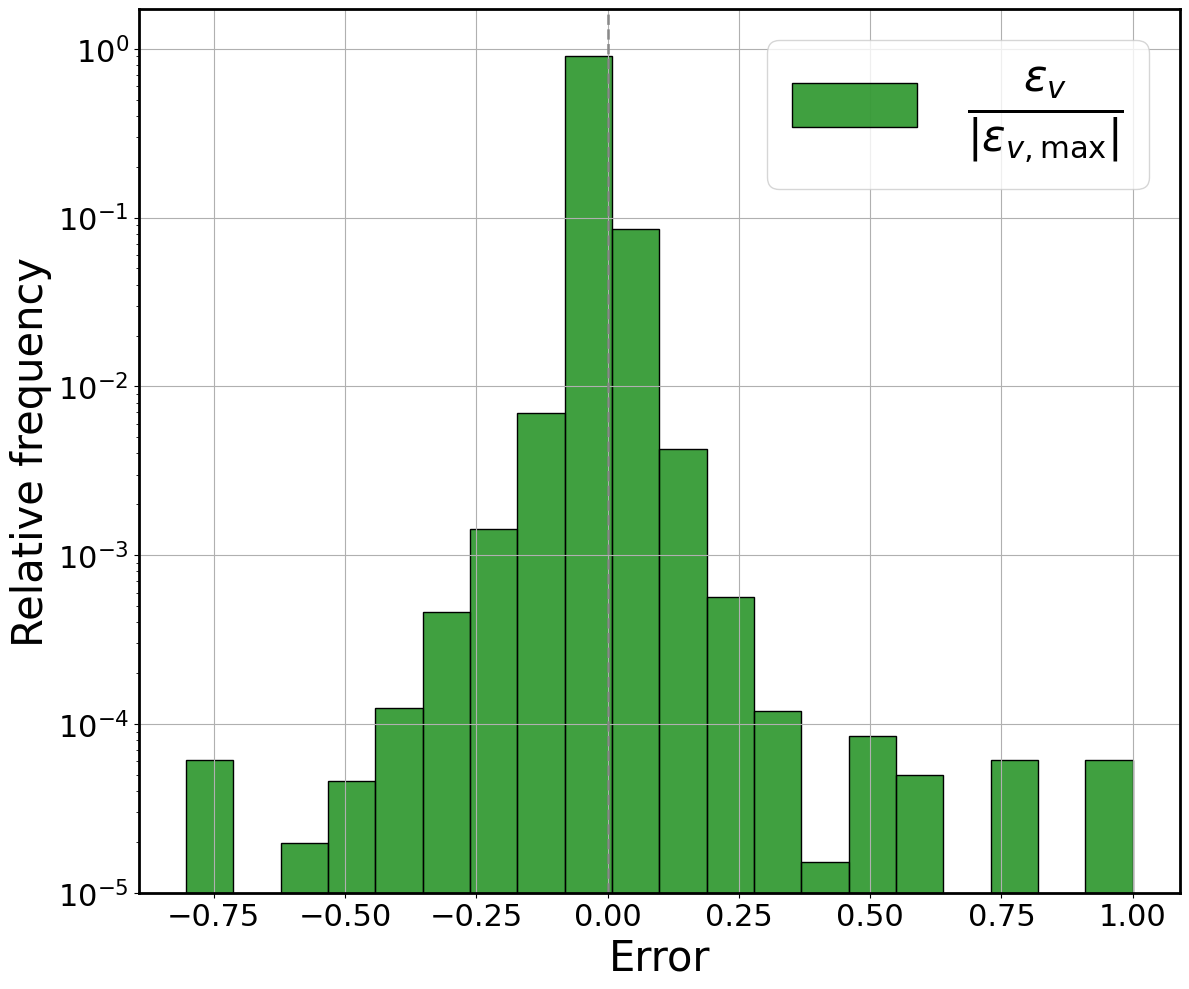}
        \caption{}
        \label{fig:N2_norm}
    \end{subfigure}
    \caption{N2 task, Same as \ref{fig:F2_relfreq} but with Re = 245 and AoA = $11^\circ$.}
    \label{fig:N2_relfreq}
\end{figure}

The normalized error distribution for the N2 test case is presented as relative frequency histograms for the streamwise and normal velocity components in Fig.~\ref{fig:N2_relfreq}. The normalized error for the streamwise velocity component (Fig.~\ref{fig:N2_st}) shows that over $90\%$ of the values concentrated near zero and a slight bias toward negative values. The normalized error for the normal velocity component also has a distribution centered around zero, with a slight bias toward negative values and a higher relative frequency compared to the streamwise component. This observation is consistent with Tab.~\ref{tab:rrmse}, where the RRMSE for the normal velocity component is lower than that of the streamwise component.

\subsection{Summary of the results obtained.}

\begin{table}[h!]
\centering
\renewcommand{\arraystretch}{1.3}
\begin{tabular}{ccc}
\hline
\textbf{ID} (Tab. \ref{tab: testcases})& \textbf{RRMSE Streamwise Velocity} ($\times 10^{-2}$) & \textbf{RRMSE Normal Velocity} ($\times 10^{-2}$) \\
\hline
FRe  & 1.92757 & 1.74942 \\
FAoA & 1.22559 & 0.67426 \\
N1   & 1.35854 & 0.75767 \\
N2   & 1.53968 & 1.15253 \\
\hline
\end{tabular}
\caption{Relative Root Mean Square Error (RRMSE) for streamwise and normal velocity components for the different test cases addressed using the proposed MoTIF.}
\label{tab:rrmse}
\end{table}
Table \ref{tab:rrmse} presents the RRMSE computed for both the streamwise and normal velocity components across the four test cases considered in this work. It can be observed that the FAoA case yields the lowest error values for both velocity components, indicating that the proposed MoTIF effectively captures variations related to changes in the angle of attack. In contrast, the FRe case exhibits the highest RRMSE values, particularly for the streamwise component, suggesting that the model encounters greater difficulty generalizing across different Reynolds numbers. For the N1 and N2 cases, the RRMSE values lie between those of FRe and FAoA, with N1 showing slightly better performance than N2 in both components. The streamwise velocity tends to exhibit higher reconstruction errors than the normal velocity, likely due to the larger magnitude variations typically present in the streamwise direction. It is worth mentioning that the RRMSE for both velocity components remains consistently below $2\%$ across all test cases, demonstrating the high accuracy and robustness of the proposed model.

\section{Conclusions}\label{conclusions}

To the best of the authors' knowledge, MoTIF is introduced for the first time in this study as a unified, modular framework that simultaneously addresses three critical tasks in fluid dynamics modeling: parametric interpolation, temporal forecasting, and resolution enhancement. MoTIF enables the generation of multi-parametric databases by interpolating across physical parameters, in this case, angle of attack (AoA) and Reynolds number (Re), while extending its capabilities to produce accurate temporal predictions and high-resolution reconstructions for both ground truth and predicted data.

The results demonstrate that MoTIF delivers accurate reconstructions in both temporal and parametric domains, as well as improved spatial resolution. Its modular architecture ensures flexibility and generalizability, making it applicable to a wide range of databases where parametric variability, spatio-temporal dynamics, and resolution limitations are key challenges.

The application of higher-order singular value decomposition (HOSVD) facilitates effective dimensionality reduction while preserving the most significant flow structures, thus enabling computationally efficient data processing and storage. Coupling HOSVD with neural networks allows for spatial resolution enhancement by reconstructing high-resolution datasets from low-resolution inputs with minimal error. Furthermore, Gaussian process regression (GPR) proves effective for interpolating new flow conditions, such as intermediate Re numbers and AoA values, ensuring the continuity and consistency of fluid dynamics databases.

Validation results confirm the accuracy and generalization capacity of the proposed framework, with reconstruction errors consistently below 2\% across all test cases. The forecasting module, based on recurrent neural networks (RNNs), accurately predicts future flow states while preserving the coherence of flow structures over time.

This hybrid deep learning multi-parametric reduced-order modeling (ROM) framework represents a significant advancement in data-driven fluid dynamics analysis, offering a scalable, flexible, and computationally efficient alternative to traditional high-fidelity simulations. Its modular design renders it adaptable to diverse fluid dynamics scenarios, extending its applicability to complex flow problems beyond the configurations examined in this work.

\section{Acknowledgements}

The authors acknowledge the MODELAIR project that has received funding from the European Union’s Horizon Europe research and innovation programme under the Marie Sklodowska-Curie grant agreement No. 101072559. S.L.C. acknowledges the ENCODING project that has received funding from the European Union’s Horizon Europe research and innovation programme under the Marie Sklodowska-Curie grant agreement No. 101072779. The results of this publication reflect only the author's view and do not necessarily reflect those of the European Union. The European Union can not be held responsible for them. The authors acknowledge the grant PLEC2022-009235 funded by MCIN/AEI/ 10.13039/501100011033 and by the European Union “NextGenerationEU”/PRTR and the grant PID2023-147790OB-I00 funded by MCIU/AEI/10.13039 /501100011033 /FEDER, UE. The authors gratefully acknowledge the Universidad Politécnica de Madrid (www.upm.es) and CeSViMa for providing computing resources on the Magerit Supercomputer.

\section{Data availability}
Data sets generated during the current study are available from the corresponding author on reasonable request.

\section*{Conflict of interest}

 The authors declare that they have no conflict of interest.

\bibliographystyle{plainnat}

\bibliography{sample.bib}



\end{document}